\documentclass[%
 reprint,
 amsmath,amssymb,
 aps,
]{revtex4-2}

\usepackage{graphicx}% Include figure files
\usepackage{dcolumn}% Align table columns on decimal point
\usepackage{bm}% bold math
\usepackage{siunitx}

\usepackage{soul}
\usepackage{xcolor}
\sethlcolor{yellow}

\begin{document}

\preprint{APS/123-QED}

\title{An Electrically Injected and Solid State Surface Acoustic Wave Phonon Laser}% Force line breaks with \\

\author{Alexander Wendt$^1$}
\author{Matthew J. Storey$^2$}%
\author{Michael Miller$^2$}
\author{Dalton Anderson$^1$}
\author{Eric Chatterjee$^1$}
\author{William Horrocks$^1$}
\author{Brandon Smith$^2$}
\author{Lisa Hackett$^2$}
\author{Matt Eichenfield$^{1,2}$}\email{eichenfield@arizona.edu}

\address{$^1$Wyant College of Optical Sciences, University of Arizona, Tucson, AZ 85719 \\
$^2$Sandia National Laboratories, Albuquerque, NM 87185 \\}
%\affiliation{College of Optical Sciences, SNL
%}%


\date{\today}% It is always \today, today,

\begin{abstract}
Surface acoustic waves (SAWs) enable a wide array of technologies including RF filters, chemical and biological sensors, acousto-optic devices, acoustic control of microfluidic flow in lab-on-a-chip systems, and quantum phononics. While numerous methods exist for generating SAWs, they each have intrinsic limitations that inhibit performance, operation at high frequencies, and use in systems constrained in size, weight, and power. Here, for the first time, we present a completely solid-state, single-chip SAW phonon laser that is comprised of a lithium niobate SAW resonator with an internal, DC electrically injected and broadband semiconductor gain medium with $<$0.15 mm$^2$ footprint. Below the threshold bias of 36 V, the device behaves as a resonant amplifier, and above it exhibits self-sustained coherent oscillation, linewidth narrowing, and high output powers. A continuous on-chip acoustic output power of up to -6.1 dBm is generated at 1 GHz with a resolution-limited linewidth of $<$77 Hz and a carrier phase noise of -57 dBc/Hz at 1 kHz offset. Through detailed modeling, we show pathways for improving these devices' performance including mHz linewidths, sub -100 dBc/Hz phase noise at 1 kHz, high power efficiency, footprints less than 550 \si{\micro\meter}$^2$ at 10 GHz, and SAW generation approaching the hundreds of GHz regime. This demonstration provides a fundamentally new approach to SAW generation, paving the way toward ultra-high-frequency SAW sources on a chip and highly miniaturized and efficient SAW-based systems that can be operated without an external RF source.
\end{abstract}
\maketitle
\section{Introduction}
From RF filters in wireless communication devices to advanced biosensing platforms, surface acoustic waves (SAWs) have become integral to a wide array of technologies. Their unique property of confining acoustic energy to the surface enables remarkably efficient signal processing \cite{morgan2010surface, hashimoto2000surface}, sensitive detection\cite{mandal2022surface, lu2025harnessing, schell2023exchange, chen2020ultrahigh}, and strong interaction with physical systems through localized strain fields. These properties make them critical for real-time detection in chemical \cite{li2023advances} and biological sensors \cite{agostini2021ultra, greco2020ultra}, while also facilitating precise manipulation of liquids and particles in microfluidic devices\cite{yang2022harmonic, ding2013surface, qin2021acoustic, zhao2025topological}, empowering applications ranging from droplet control to cell sorting \cite{li2015acoustic}. In addition, SAWs have emerged as a tool for qubit state manipulation in quantum processors \cite{pirkkalainen2013hybrid, gustafsson2014propagating, whiteley2019spin, maity2020coherent}, transduction for quantum networking \cite{manenti2017circuit, schutz2017universal, decrescent2022large, zhou2024electrically}, and even as the basis for phononic quantum computing \cite{manenti2017circuit, arrangoiz2019resolving, sletten2019resolving, qiao2023splitting, qiao2025acoustic}. Their strong interaction with electromagnetic radiation has been used in Brillouin-based optical devices \cite{shin2015control, zhou2024nonreciprocal, freedman2025gigahertz}, and as sources for photoelastic interactions in acousto-optic modulation \cite{shao2019microwave, hassanien2021efficient, kittlaus2021electrically}, with applications including LiDAR \cite{li2023frequency, lin2024optical}, optical computing \cite{zhao2022enabling}, and optical control of quantum computers \cite{neuman2021phononic}.

SAWs are most commonly generated by transducing an RF electrical signal using an interdigital transducer (IDT) \cite{morgan2010surface}. While transduction is essential for RF filtering applications - where ultimately the electrical signal is of interest - many other applications such as those outlined above function independently of the acoustic wave generation mechanism. These applications, when enabled by an external source with IDTs, then inherent the limitations intrinsic to this architecture, which are as follows. First, IDTs rely on impedance-matched periodic electrode structures with significantly sub-wavelength metal thickness for efficient transduction. At frequencies in the tens of GHz, this metal thickness becomes prohibitively thin, introducing both parasitic resistance and unwanted acoustic reflections, ultimately limiting their efficiency and scalability. Second, self-contained systems require at least one other electronic chip for the RF source, increasing the size, weight, and power (SWaP). Alternatively, electronic SAW oscillators, which incorporate a SAW resonator directly into a regenerative feedback loop via transduction from IDTs, could offer a means of SAW generation without an entirely external source. These devices have demonstrated impressive performance \cite{lu2012novel, bahr201516, xi2025low}, and could be modified to couple out some of the acoustic signal as a source of coherent SAWs. However, as the amplification is in the electrical domain, IDTs must be used to transduce to and from the acoustic domain, limiting the devices in ways mentioned above. 

\begin{figure*}
  \centering
  \includegraphics[width=\linewidth]{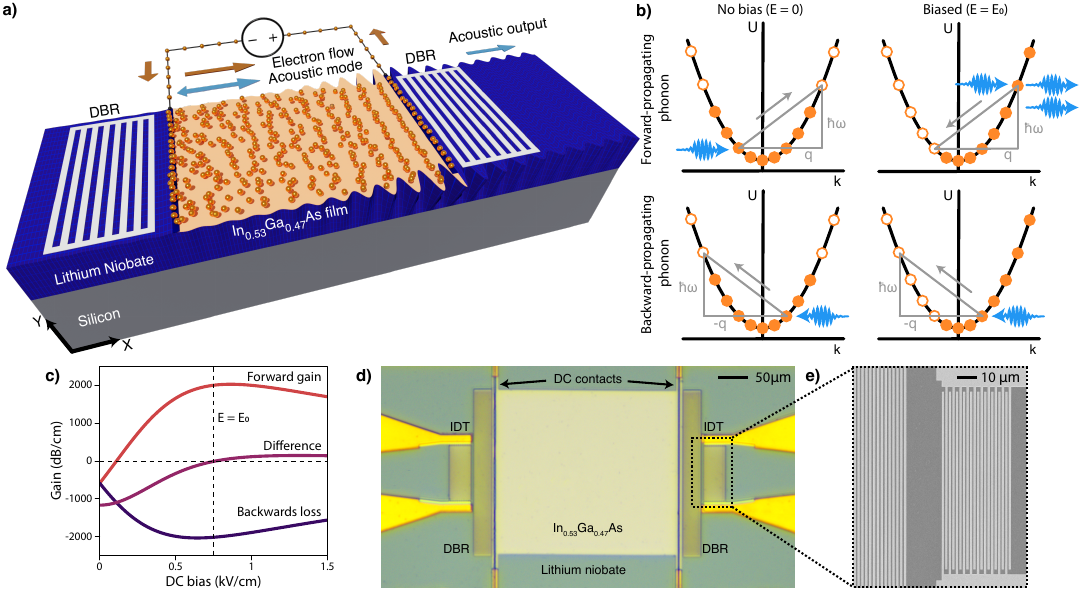}
  \caption{
    \textbf{The phonon laser.}
    \textbf{a)} Schematic illustration of the SAW-PL. An acoustoelectric amplifier formed from heterogeneously integrated In$_{0.53}$Ga$_{0.47}$As on lithium niobate is incorporated inside an acoustic resonator. The amplifier is able to overcome the intrinsic losses of the resonator, resulting in a coherent lasing process.
    \textbf{b)} The acoustoelectric effect may be thought of as an absorption/emission process between the momentum states of electrons in the semiconductor. The presence of a DC bias gives the electrons a nonzero average momentum, shifting the occupied states (filled circles) to the right. This creates a population inversion for forward propagating phonons, while still absorbing backward-propagating phonons.
    \textbf{c)} Theoretical gain curve from classical theory as a function of drift field. At a bias $E_0$, the forward gain may overcome the backward loss, allowing net round-trip amplification.
    \textbf{d)} Micrograph of a device. Device consists of two acoustic DBRs with an acoustoelectric heterostructure interposed between them. IDTs on either side used for readout, and to probe the device when biased below threshold.
    \textbf{e)} Scanning electron microscope image detailing the IDT and DBR found on one side of the device.
    }
    \label{fig: 1}
\end{figure*}

An alternative method for generating SAWs on a chip is to use an optomechanical phonon laser, where radiation pressure is used to parametrically amplify acoustic fields in mechanical resonators \cite{vahala2009phonon, grudinin2010phonon, zhang2018phonon, pettit2019optical}. When a resonator is pumped with blue-detuned light, phonon emission is amplified in a process analogous to optical lasing. However, these devices require simultaneous photonic and phononic engineering of the system. They also require a precisely tuned laser source. These make the system substantially complex and limit their use in low-SWaP applications. Similarly, other optomechanical processes with multiple lasers have been used to generate coherent SAWs \cite{iyer2024coherent}. This promising technique enables phonon generation in systems that are not optically transparent or piezoelectric. However, this technique requires multiple optical fields with precise frequency and angular separation, hindering or even precluding their use in many integrated systems.

The acoustoelectric effect has been thought to provide a pathway to direct acoustic oscillation since its discovery, and even was successful in generating weak bulk wave oscillations \cite{okada1964continuous, maines1968gigahertz, maines1970current}. These early demonstrations, however, were unable to effectively couple to SAWs, required strong biases and complicated illumination schemes, were limited to piezoelectric semiconducting materials such as cadmium sulfide and zinc oxide, and only demonstrated current oscillations as a result of the acoustoelectric effect. Since these initial demonstrations, the advancement of semiconductor growth techniques and nano-scale engineering has re-ignited interest in the acoustoelectric effect. While it has been used to modestly improve the quality factor of MEMS resonators \cite{gokhale2014phonon, mansoorzare2019acoustoelectric}, these demonstrations have been far from the regime of overcoming in the intrinsic losses of the resonator. By heterogeneously integrating epitaxial thin films of high-mobility semiconductors onto strongly piezoelectric substrates, a strong acoustoelectric interaction can take place. This approach has demonstrated impressive performance for amplification \cite{hackett2023non}, non-reciprocal transmission \cite{hackett2021towards}, and phononic mixing \cite{hackett2024giant}. 

Here we report an entirely self-contained, solid state, and electrically injected surface acoustic wave phonon laser (SAW-PL), utilizing an acoustoelectric amplifier as a gain medium. By heterogeneously integrating an amplifier within a SAW resonator, the thermally-occupied modes of the resonator achieve round trip gain to undergo sustained phononic lasing. The paper is organized as follows: First, we describe the device architecture, and show how a DC bias alters the electronic momentum states to create a population inversion for phonons. Then, we characterize the device when biased below threshold, where it acts as a resonant amplifier. Next, we demonstrate a clear change in behavior above threshold, with the linewidth narrowing from the MHz regime to under 77 Hz, and measure the power, linewidth, and phase noise. We discuss an alternative architecture that we show could improve the linewidth by two orders of magnitude and the phase noise by almost 45 dB. Finally, we propose several near term applications of this device, and show through  modeling the potential for this technology to be scaled to tens of GHz. This work marks a critical milestone in the development of acoustic-based RF systems, paving the way for the integration of SAW devices at the system level.

\section{Operating principle}
The SAW phonon laser (SAW-PL), illustrated in Fig. 1a, consists of an acoustoelectric amplifier heterostructure fabricated inside an acoustic resonator. The device studied in this manuscript uses two acoustic mirrors to form a Fabry-Perot cavity, but it is also possible to use traveling-wave or ring cavity, discussed extensively below. The amplifier heterostructure consists of a high-mobility semiconducting film heterogeneously integrated on a piezoelectric substrate. The evanescent electric field of the phonons extends into the semiconducting layer, enabling strong coupling between the acoustic field and the charge carriers \cite{kino1971normal}. The device operates analogously to a semiconductor optical laser, where the DC electrically injected amplifier heterostructure provides gain for the thermally-populated phonon modes of the resonator. When the gain medium is able to provide enough gain to overcome the round-trip losses for any mode, the system will undergo self-oscillation, leading to coherent emission at the frequency of the dominant mode.

Acoustoelectric amplification may be thought of as a stimulated emission process, much like in lasers \cite{pippard1963acoustic}, but between the electronic momentum states in the semiconductor rather than energy levels in a two-level system. To illustrate this, consider a 1-D free electron gas, which has dispersion relationship $E = \frac{\hbar^2k^2}{2m_e}$, where $m_e$ is the electron mass. A phonon propagating in a medium with acoustic velocity $v_a$ with wavenumber $q$ will have energy $\hbar\omega = \hbar v_aq$. An electron may interact with this phonon if and only if the interaction conserves energy and momentum, which is the case when $\hbar \bar{k} = m_e v_a$, where $\bar{k} = \frac{k_i + k_f}{2}$ is the average electron wave number of the initial and final states. In this 1-D picture, there is exactly one unique pair of electronic states a forward propagating phonon can interact with, and a different unique pair of states a backward propagating phonon can interact with. These phonons will have momenta +q and -q, respectively. In the absence of bias, the electrons occupy the lowest energy levels, and may absorb phonons propagating in either direction, shown in the first column of Fig. 1b.

When a DC bias, $E$, is applied, the average momentum of the electrons becomes $\hbar k_0 = m_e \mu E$, where $\mu$ is the electron mobility. This effectively shifts the occupied electronic states to the right in k-space, shown in the second column of Fig. \ref{fig: 1}b. For forward propagating phonons, there is a bias $E_0$ at which the higher energy state in its unique pair becomes populated, while the lower energy state is unpopulated, enabling both stimulated and spontaneous emission for that phonon mode. For backward phonons, however, this same bias shifts the occupied states away from the high-energy level of its unique pair, and will continue to absorb phonons. Thus, one can view this bias as creating a population inversion for forward phonons, but not for backward propagating phonons. In higher dimensions, there are more pairs of electronic states that satisfy the conservation laws, as only the projection of the electronic momentum must be in the direction of phonon propagation. This gives rise to population statistics, modifying the rates of these processes. This is discussed in depth in Supplemental note A, along with considerations for non-zero temperature, and energy and momentum broadening effects.
\begin{figure*}[htbp]
  \centering
  \includegraphics[width=\linewidth]{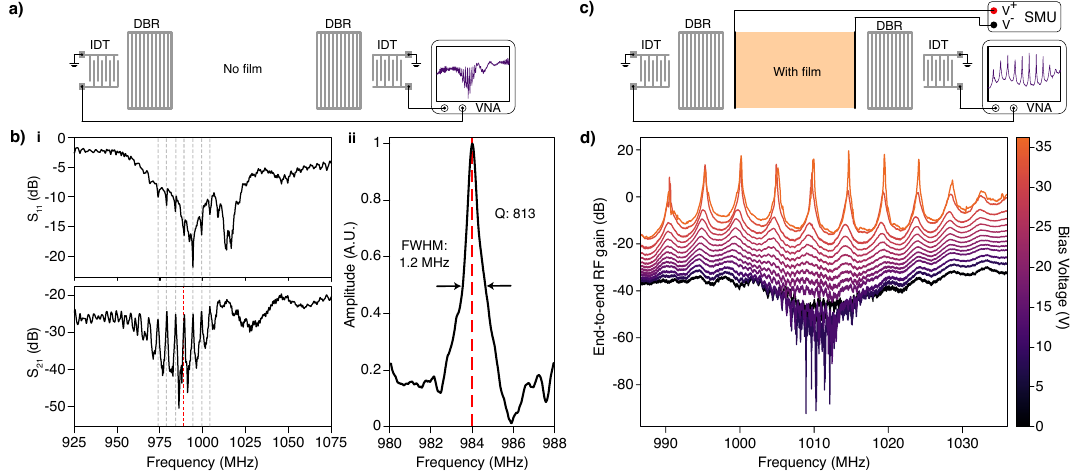}
  \caption{
    \textbf{Passive resonator and resonant amplifier characterization.}
    \textbf{a)} Passive resonator experimental setup. Frequency sweeps are performed with a vector network analyzer (VNA).
    \textbf{b)} Characterization of a passive resonator. \textbf{i} S-parameter data and \textbf{ii} zoomed in peak with FWHM measurement. Dips in the reflection (S$_{11}$) spectrum correlate with peaks in the transmission (S$_{21}$) spectrum, indicating resonator resonances.
    \textbf{c)} Resonant amplifier experimental setup. Similar to the passive resonator measurement, but using a source measurement unit (SMU) to apply a constant bias across the semiconductor found in the SAW-PL.
    \textbf{d)} Resonant amplifier characterization when devices are biased below threshold. Peak end-to-end gain of 19.6 dB, the largest acoustoelectric amplification to date.
  }
  \label{fig: 2}
\end{figure*}

This quantum picture allows one to think of our phonon laser analogously to a semiconductor laser. It can also be used to derive the change in the phonon propagation constant as $\beta' = \beta_a +i\alpha_{ae} + \beta_{ae}$, where $\beta_a$ is the unperturbed acoustic wavenumber, and $\alpha_{ae}$ and $\beta_{ae}$ are the gain and velocity-dependent propagation constant modifications to the wavenumber. However, due to the effects of finite temperature and collisional broadening, analytical expressions for $\alpha_{ae}$ and $\beta_{ae}$ are difficult to derive. Instead, we may use a drift diffusion model that yields classical expressions for the acoustoelectric effect that have proven to be exceptionally accurate \cite{hackett2021towards, mansoorzare2019acoustoelectric, hackett2023non, hackett2024giant}. These expressions are given by

\begin{align}
    \alpha_{ae} = \frac{1}{2} k^2_{\text{eff}}\beta_a\frac{(v_d/v_a - 1)(\omega_c/\omega)}{(v_d/v_a - 1)^2 + (\frac{R\omega_c}{\omega} + H)^2}
    \label{eq: ae gain}
\end{align}
\begin{align}
    \beta_{ae} = \frac{1}{2} k^2_{\text{eff}}\beta_a \frac{(R\omega_c/\omega + H)(\omega_c/\omega)}{(v_d/v_a - 1)^2 + (\frac{R\omega_c}{\omega} + H)^2}
    \label{eq: ae prop}
\end{align}
where $k^2_{\text{eff}}$ is the effective electromechanical coupling coefficient, $v_d = \mu E$ is the electron drift velocity, $R$ is the space-charge reduction factor, $\omega_c = \mu|\rho_0|/\epsilon_s$ is the dielectric relaxation frequency, and $H$ is a diffusion term that takes into account the diffusivity of the medium \cite{kino1971normal, coldren1972monolithic} (additional details in supplemental note B).

Fig. 1c depicts the gain curve as a function of bias. As the bias increases, the forward propagating field receives gain, while the backward propagating field continues to be lossy. However, the rate of these processes are not symmetric about zero bias. Consequently, the bias at which the loss peaks is lower than the bias at which the gain peaks. Thus, there is a bias at which forward-propagating phonons' gain is large enough to overcome the backward-propagating phonons' loss. Furthermore, if this net gain overcomes the other losses present in the resonator, then the system will have zero net loss, and will self-oscillate. An intuitive understanding of the shape of the gain curve in the context of stimulated emission and absorption is included in supplemental note A.

A labeled micrograph of the device studied in this manuscript is shown in Fig. 1d. The IDTs that surround the SAW-PL have two purposes. First, to both source and collect acoustic waves to probe the resonator below threshold, where it acts as a resonant amplifier. Second, above threshold when the device self oscillates, we use one IDT to collect the coherent acoustic radiation generated from the SAW-PL. The specific SAW-PL implementation studied in this work utilizes metallic distributed Bragg reflectors (DBRs) as the acoustic mirrors, which are formed by shorting a series of periodically spaced aluminum lines, shown in greater detail in the SEM depicted in Fig. 1e. While these structures are superficially similar to IDTs in their use of periodic electrodes, they do not suffer from the same limitations. In IDTs, electrode thickness must be carefully optimized: if too thick, the electrodes excessively mass load the surface and introduce spurious reflections; if too thin, parasitic resistance degrades transduction efficiency. As DBRs serve only as passive acoustic reflectors, they are not subject to these competing constraints. The amplifier heterostructure consists of a high-mobility, $50$ nm thick indium gallium arsenide (In$_{0.53}$Ga$_{0.47}$As) semiconducting film heterogeneously integrated on a 5-\si{\micro\meter}-thick, Y-cut lithium niobate (LN) piezoelectric layer on a silicon substrate (see methods for details). The device hosts a SAW mode with primarily quasi-shear horizontal (q-SH0) polarization and high electromechanical coupling coefficient ($k^2 >13\%$) that propagates in the material X-direction around 1 GHz \cite{hackett2023non}. The total footprint of the device, including the mirrors and amplifier, was 460 \si{\micro\meter} x 300 \si{\micro\meter}, with a full list of dimensions in supplemental note C.

\begin{figure*}[htbp]
  \centering
  \includegraphics[width=\linewidth]{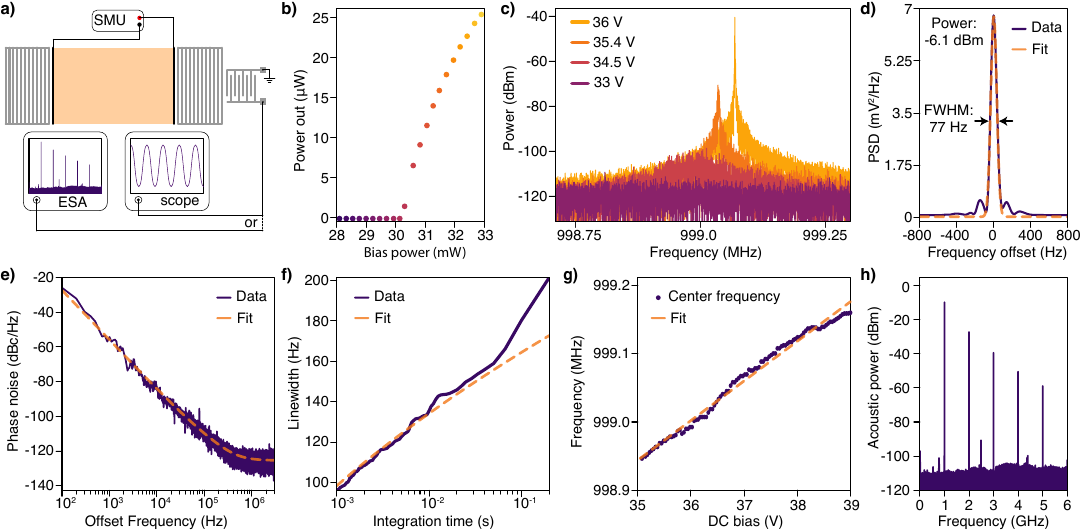}
  \caption{
    \textbf{Phonon lasing characterization.}
    \textbf{a)} Experimental setup, with the output of the SAW-PL sent to an oscilloscope (scope) for figs. (b-g), or a spectrum analyzer (ESA) for h.
    \textbf{b)} Output power as function of input power, demonstrating a clear threshold.
    \textbf{c)} Linewidth narrowing as threshold is achieved. The linewidth narrows from about 1.2 MHz to $<$100 Hz. 
    \textbf{d)} Characterization of linewidth and output power, finding a linewidth of 77 Hz and -6.1 dBm of acoustic power.
    \textbf{e)} Phase noise with fit using Leeson's equation, using the calculated acoustoelectric loss-limited Q of 96 and noise factor of 2.
    \textbf{f)} Calculated linewidth from frequency noise as a function of integration time.
    \textbf{g)} Center frequency tuning of $55.4\pm0.6$ kHz/V.
    \textbf{h)} Full output spectrum. Harmonics are perfect multiples of the fundamental, suggesting nonlinear mixing.
  }
  \label{fig: 3}
\end{figure*}

\section{Device characterization}
We begin by discussing the operation of the device biased below threshold, where it behaves as a resonant amplifier, shown in Fig. 2. We first characterized a passive resonator fabricated on the same chip that does not contain the epitaxial semiconductor layer using frequency sweeps from a vector network analyzer (VNA), shown in Fig. 2a. The S-parameters of one such resonator (more details in the methods section), are shown in Fig. 2bi. There are resonance peaks in the transmission spectrum (S$_{21}$) that correspond to dips in the reflection spectrum (S$_{11}$) with spacings of $\Delta\nu_{\text{FSR}} = 5.1\pm0.04$ MHz. Figure 2bii shows the transmission peak at 984 MHz in greater detail, which we find has a full-width half-max (FWHM) of $ \Delta\nu=1.20\pm0.01$ MHz, corresponding to a quality factor of 813.

SAW-PLs were fabricated with IDTs on either side in order to probe the devices with a coherent signal to characterize their behavior as resonant amplifiers. An experimental diagram is shown in Fig. 2c, where a source measurement unit (SMU) is used to bias the semiconductor, while VNA frequency sweeps measure the device response. A device with the same dimensions as the passive resonator but with an amplifier that filled the resonator was fabricated on the same chip and characterized, shown in Fig. 2c. We see that, at low biases, the transmission is extremely lossy when compared to a passive resonator. This is expected as the acoustoelectric effect is lossy at zero bias, as previously described. Near threshold, the device achieves net acoustic amplification, with up to 19.6 dB of end-to-end RF gain, the largest measured in an acoustoelectric device to date. By estimating the insertion losses of the IDTs (details in supplemental note D) and the propagation loss, we calculate there is 36.3 dB of acoustic gain on chip when biased just below threshold at 36 V. At this bias, we fit the Q factor of the peak near 1018 MHz using a Lorentzian line shape to be 1688, a 108$\%$ increase over the passive resonator Q factor, to our knowledge, the largest acoustoelectric quality factor enhancement reported to date.

When biased above threshold, the SAW-PL demonstrates behavior consistent with self-oscillation. The same device that was tested as a resonant amplifier was characterized in this regime, shown in Fig. 3. To do so, only one IDT is needed to transduce the acoustic output of the device, which is subsequently sent to an oscilloscope (Figs. 3b-f) or a spectrum analyzer (Fig. 3g), shown in Fig 3a. Details on data acquisition and processing are in the Methods section. A clear threshold power is shown in Fig. 3b, a signature of lasing. As the power increases through threshold, the linewidth narrows, shown in Fig. 3c. At biases just below threshold, a broad resonance appears, with a linewidth of about 1 MHz and a peak power of -100 dBm. As the voltage is swept through threshold, the power increases to almost -40 dBm. At a slightly higher bias of 33.9 mW, we integrate over a 20 ms section of a 200 ms long set of data to measure the output power and linewidth, shown in Fig. 2d. The output power was -6.1 dBm, giving a power efficiency of $0.7\%$. The linewidth was 77 Hz when fit with a Gaussian, limited by the of the length of the integration time. Integrating for longer periods of time indicate there is a slow, random walk-like jitter to the center frequency that inhibits a more precise measurement which we attribute to environmental noise. A detailed analysis of the linewidth, including calculations of the Schawlow-Townes linewidth, Henry enhancement factor, and analysis of several noise sources is in supplemental note E. 

Next, we calculated the SAW-PL's phase noise using the same dataset, depicted in Fig. 3e (see Methods for more details). By modeling our device as an electronic oscillator circuit, we can use Leeson's formula to compare it's performance to the theoretical performance. The phase noise $L$ in dBc/Hz is
\begin{equation}
    L(f_m) = 10\log_{10}\bigg[\frac{Fk_BT}{2P_0}\bigg(\bigg(\frac{f_0}{2Q_{\text{eff}}f_m}+ 1 \bigg)^2 \bigg(\frac{f_c}{f_m}+1 \bigg) \bigg]
    \label{eq: Leeson's}
\end{equation}
where $f_0$ is the center frequency, $f_m$ is the offset from the center frequency, $f_c$ is the corner frequency, F is the noise factor, $k_B$ is the Boltzmann constant, T is the temperature, and $P_0$ is the acoustic power incident on the internal amplifier \cite{rhea1990oscillator}. By simultaneously measuring the signal from the front and back ports of the device, we estimate the acoustoelectric loss to be -17.8 dB, resulting in an effective Q factor of 96. Using this tandem with the calculated loss from the Q factor, we estimate $P_0 = 2.5$ \si{\micro\watt}. Finally, we set the noise factor $F = 2$, as characterized by previous work \cite{hackett2023non}. Using this set of parameters, we fit the flicker frequency $f_c = 50$ kHz, and find good agreement with what is experimentally observed. By calculating the frequency noise and integrating over 1 ms, we find a linewidth of about 92 Hz, in good agreement with the FFT method outlined above. Additional details of these calculations are in supplemental note E. A straightforward way of improving the phase noise by more than 27 dB is discussed in depth in the discussion section.

The linewidth $\Delta\nu$ of an oscillator may be estimated by integrating the power spectral density of frequency noise $S_f(f) = 2f^2 10^{L(f) / 10}$ with respect to frequency and taking the square root:
\begin{align}
    \Delta \nu(t) \approx \sqrt{\int_{1/t}^{f_{\text{max}}} S_f(f) df}
\end{align}
where $t$ is the integration time and $f_{\text{max}}$ is the maximum frequency, chosen to be 10 kHz to avoid contributions from the oscilloscope's jitter (details in supplemental note E). We calculate this numerical from the phase noise data, and from the Leeson's formula, shown in Fig. 3f. We observe similar linewidths as the Fourier transform method, and find good agreement between the data and fit until about 100 ms, after which they diverge. This is likely due to low frequency jitter, which only impact the device on timescales $<$10 Hz.

The acoustoelectric effect modifies the propagation constant, which will in turn modify the round-trip phase condition, changing the resonant frequency of the SAW-PL. By ramping the bias field over time, we see this present in Fig. 3g, which shows the center frequency as a function of bias. From the slope of this graph, we extract a tuning rate of $55.4\pm0.6$ \si{\kilo\hertz/\volt} for this device. This is similar to what one would expect by calculating the velocity shift due to the acoustoelectric effect, with details in supplemental note F.

The full output power spectrum is shown in Fig. 3h, revealing harmonics of the fundamental. These are unlikely to be independent oscillations, given that they are exact multiples of the fundamental frequency. Previous work has shown this platform has strong acoustic nonlinearities \cite{hackett2024giant}, making it likely that these harmonics are due to a nonlinear mixing process. This is further supported by the fact that FEM models indicate that even though the DBRs are reflective at these higher frequencies, the reflection coefficients are lower, the propagation loss is higher, and the electromechanical coupling coefficients are lower (more details in supplemental note C).

\begin{figure*}[htbp]
  \centering
  \includegraphics[width=\linewidth]{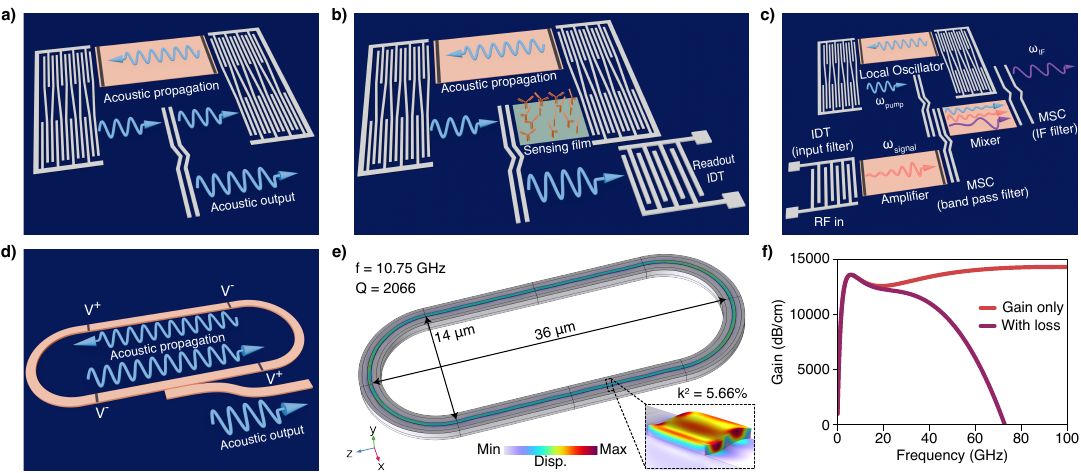}
  \caption{
    \textbf{Improved architectures for near term applications and high frequency generation.}
    \textbf{a)} Ring resonator example. Such a topology would eliminate backward propagation and improve Q factors and phase noise.
    \textbf{b)} SAW-PL-based sensor. By incorporating a thin film into a device, small velocity perturbations from external stimuli, in this case antibodies, result in a detectable resonance shift read out by an IDT.
    \textbf{c)} Compact RF signal processor, consisting of an RF input, which is subsequently amplified, then mixed with a local oscillator generated from an SAW-PL to produce a low frequency signal $\omega_{\text{out}}$
    \textbf{d)} SAW-PLs may be made using waveguide-based resonators to generate high frequency SAWs.
    \textbf{e)} FEM model of a racetrack resonator made from the epitaxial semiconductor layer, supporting a high Q mode.
    \textbf{f)} Theoretical scaling of SAW-PLs. Using state of the art resonators, one can reasonably estimate that self-oscillation may occur at frequencies above 70 GHz.
    }
  \label{fig: 4}
\end{figure*}

\section{Discussion and Outlook}
Here we have demonstrated the first electrically injected surface acoustic wave phonon laser, which has a small footprint of 0.138 mm$^2$, narrow intrinsic linewidth of at most 77 Hz, and strong acoustic output power of -6.1 dBm. While we have experimentally demonstrated it is possible to overcome the acoustoelectric losses from reverse propagation inherent to a Fabry-Perot architecture, the backward loss is the dominant mechanism that limits both the threshold bias and power, as well as the linewidth and phase noise. A ring oscillator architecture would eliminate the backward propagation, substantially improving all these metrics. An illustration of one implementation is depicted in Fig. \ref{fig: 4}a, which utilizes reversing multi-strip couplers \cite{danicki1993reversing} to reverse the acoustic wave propagation direction and separate it spatially from the incident wave. These types of resonators work in a similar fashion to the DBRs used here and should not substantially impact the loss. Operating at the same frequency and in the same material system, the resonator Q factor would return to the passive value of 813, about an order of magnitude improvement of the limited value of 96. This would reduce the threshold bias from 1.09 kV/cm to approximately 0.13 kV/cm, in turn reducing the power dissipated across the device by almost two orders of magnitude. Additionally, as seen in Leeson's formula, this Q factor improvement should improve the phase noise by 18.6 dB, and should also improve the circulating power $P_0$ by a factor of about 8.5, for a net improvement of 27.8 dB. By further improving the Q factor to 2800, achieved in other lithium niobate based SAW oscillators \cite{xi2025low}, this improvement would further increase to a total of 44 dB lower than it is present, assuming external noise sources are sufficiently suppressed.

Incorporating materials sensitive to external stimuli inside the resonator transforms SAW-PLs into ultra-low SWaP sensors with high sensitivity. For example, magnetostrictive materials such as FeGaB enable magnetic field sensing, while specialized binding films allow for biological and gas detection. Fig. 4b illustrates one potential implementation, where biological analytes bind to a sensing film, altering the resonant frequency, which is transduced by an IDT so that it may be detected. Similarly, SAW-PLs may be incorporated into opto-mechanical circuits as a source of strong, DC-powered coherent SAWs, providing precise frequency modulation and deflection of optical beams in low-SWaP systems.

This work is a step closer to the realization of an all-acoustic RF front end fabricated on a single chip. Current RF signal processors rely on system-level integration of multiple technologies using different material platforms via interposers to create multi-chip modules, limiting miniaturization and performance due to multiple transduction steps. An all-acoustic RF front end would be ultra-compact, and many of the components have been demonstrated using acoustoelectrics, including amplifiers, circulators, and mixers \cite{hackett2021towards, hackett2023non, hackett2024giant}. The SAW-PL can be used as a local oscillator, a crucial piece that had yet to be demonstrated until now. Figure \ref{fig: 4}c illustrates an ultra-compact RF signal processing device, replacing six components (input filter, low-noise amplifier, interject filter, local oscillator, IF mixer, and IF filter) of an RF front-end receive chain - currently made with multiple technologies - with all-acoustic counterparts. Incoming RF signals at $\omega_{\text{signal}}$ are filtered by a narrow-bandwidth IDT, and subsequently amplified by an acoustoelectric amplifier. A multistrip coupler \cite{morgan2010surface} then filters and couples the output to a track with the output of an SAW-PL, which produces a strong pump at frequency $\omega_{\text{pump}}$. An acoustoelectric mixer then down-converts the signal to a frequency $\omega_{\text{IF}}$, subsequently filtered by another multi-strip coupler. Using parameters similar to the system presented here, we estimate that such a device operating around 5 GHz could have a total footprint of 80 \si{\micro\meter} x 110 \si{\micro\meter}, with an local oscillator that dissipates 16.7 \si{\micro\watt} of power with a threshold voltage of 1.15 V.

SAW-PLs may serve as a source of ultra-high frequency SAWs in SWaP-constrained systems. However, the periodic electrode structures required for both DBRs and reversing multi-strip couplers becoming increasingly difficult to fabricate for high frequency SAWs. Instead, a waveguided structure, similar to an optical ring resonator, may be used as a high frequency acoustic resonator. One near-term approach using the Y-cut LN platform presented in this article could be to create a waveguide from the epitaxial layer itself, illustrated in \ref{fig: 4}e. Figure \ref{fig: 4}g depicts an FEM model of one such implementation with a bending-loss-limited Q factor of 2066 at approximately 10.75 GHz with a footprint of 36 \si{\micro\meter} x 14 \si{\micro\meter}. Using the model's predicted $k^2$ of $5.66\%$, we predict a threshold bias field of 0.49 kV/cm, and estimate it would dissipate 77 \si{\micro \watt} of power. 

To scale to ultra-high frequencies in the tens of GHz, the propagation loss of the acoustic wave must be considered, as it typically scales as $f^2$ \cite{bajak1981attenuation}. A potential material platform that supports exceptionally high $k^2$ modes of $>$40$\%$ and low losses at high frequencies are X-cut LN plates. Recent work has experimentally demonstrated high-Q acoustic resonators past 100 GHz \cite{xie2024high}. Assuming that resonators with similar F-Q products may be fabricated at lower frequencies, we can plot the net acoustoelectric gain as a function of frequency, shown in Fig. \ref{fig: 4}f. We find round-trip gain is achievable past 70 GHz for a ring resonator topology for a dc bias of 2 kV/cm. 

Finally, The heterostructure discussed here is suitable for room temperature experiments, but will fail at cryogenic temperatures due to carrier freeze out. Instead, a two dimensional electron gas (2DEG) heterostructure may be considered. The ultra high mobilities found in 2DEGs would result in SAW-PLs with low bias gain, vastly reducing the power dissipation. Additionally, phononic loss is much lower at these low temperatures \cite{shao2019phononic}, further reducing the threshold, and improving the linewidths and phase noise. Matching the Q factor of previous work could yield a linewidth of 23 mHz with an integration time of 1 s. These characteristics make them an ideal candidate as a pump for numerous quantum applications, including transduction, parametric amplification, and qubit manipulation. By harnessing the strong nonlinearities predicted in these 2DEG-piezoelectric platforms \cite{chatterjee2024ab}, exotic quantum acoustic states such as phononic squeezed states or entangled phonon pairs may be created and studied.

We have demonstrated, to our knowledge, the first solid state electrically injected surface acoustic wave phonon laser by heterogeneously integrating an In$_{0.53}$Ga$_{0.47}$As film inside a piezoelectric SAW resonator. The device exhibits lasing, with a definitive threshold power of approximately 30.3 mW and linewidth narrowing to under 100 Hz with an acoustic output power of -6.1 dBm. By probing the device with a weak signal, we have demonstrated these devices act as resonant amplifiers when biased below threshold and provide the largest end-to-end RF gain measured in an acoustoelectric device to date. Using a combination of FEM and detailed analytical modeling, we have shown these devices can be scaled in the near-term to much higher frequencies and can provide a pathway to the efficient generation of ultra-high frequency acoustic waves for SWaP-constrained systems. There are several immediate applications of this technology, including a multitude of low-SWaP sensors, ultra-compact acousto-optic modulators, and as local oscillators for an all-acoustic RF front-end. The SAW-PL marks a pivotal advancement in surface acoustic wave technology, demonstrating the first fully integrated, DC-powered acoustic oscillator in a small footprint. This technology provides a route to the development of a new generation of ultra-compact, ultra-coherent, and ultra-high frequency surface acoustic wave based devices.

\bibliographystyle{ieeetr}
\bibliography{apssamp}

\begin{thebibliography}{10}

\bibitem{morgan2010surface}
D.~Morgan, {\em Surface acoustic wave filters: With applications to electronic communications and signal processing}.
\newblock Academic Press, 2010.

\bibitem{hashimoto2000surface}
K.-y. Hashimoto and K.-Y. Hashimoto, {\em Surface acoustic wave devices in telecommunications}, vol.~116.
\newblock Springer, 2000.

\bibitem{mandal2022surface}
D.~Mandal and S.~Banerjee, ``Surface acoustic wave (saw) sensors: Physics, materials, and applications,'' {\em Sensors}, vol.~22, no.~3, p.~820, 2022.

\bibitem{lu2025harnessing}
X.~Lu, Y.~Yuan, F.~Chen, X.~Hou, Y.~Guo, L.~Reindl, Y.~Fu, W.~Luo, and D.~Zhao, ``Harnessing exceptional points for ultrahigh sensitive acoustic wave sensing,'' {\em Microsystems \& Nanoengineering}, vol.~11, no.~1, pp.~1--10, 2025.

\bibitem{schell2023exchange}
V.~Schell, E.~Spetzler, N.~Wolff, L.~Bumke, L.~Kienle, J.~McCord, E.~Quandt, and D.~Meyners, ``Exchange biased surface acoustic wave magnetic field sensors,'' {\em Scientific Reports}, vol.~13, no.~1, p.~8446, 2023.

\bibitem{chen2020ultrahigh}
Z.~Chen, J.~Zhou, H.~Tang, Y.~Liu, Y.~Shen, X.~Yin, J.~Zheng, H.~Zhang, J.~Wu, X.~Shi, {\em et~al.}, ``Ultrahigh-frequency surface acoustic wave sensors with giant mass-loading effects on electrodes,'' {\em ACS sensors}, vol.~5, no.~6, pp.~1657--1664, 2020.

\bibitem{li2023advances}
X.~Li, W.~Sun, W.~Fu, H.~Lv, X.~Zu, Y.~Guo, D.~Gibson, and Y.-Q. Fu, ``Advances in sensing mechanisms and micro/nanostructured sensing layers for surface acoustic wave-based gas sensors,'' {\em Journal of Materials Chemistry A}, vol.~11, no.~17, pp.~9216--9238, 2023.

\bibitem{agostini2021ultra}
M.~Agostini and M.~Cecchini, ``Ultra-high-frequency (uhf) surface-acoustic-wave (saw) microfluidics and biosensors,'' {\em Nanotechnology}, vol.~32, no.~31, p.~312001, 2021.

\bibitem{greco2020ultra}
G.~Greco, M.~Agostini, and M.~Cecchini, ``Ultra-high-frequency love surface acoustic wave device for real-time sensing applications,'' {\em Ieee Access}, vol.~8, pp.~112507--112514, 2020.

\bibitem{yang2022harmonic}
S.~Yang, Z.~Tian, Z.~Wang, J.~Rufo, P.~Li, J.~Mai, J.~Xia, H.~Bachman, P.-H. Huang, M.~Wu, {\em et~al.}, ``Harmonic acoustics for dynamic and selective particle manipulation,'' {\em Nature materials}, vol.~21, no.~5, pp.~540--546, 2022.

\bibitem{ding2013surface}
X.~Ding, P.~Li, S.-C.~S. Lin, Z.~S. Stratton, N.~Nama, F.~Guo, D.~Slotcavage, X.~Mao, J.~Shi, F.~Costanzo, {\em et~al.}, ``Surface acoustic wave microfluidics,'' {\em Lab on a Chip}, vol.~13, no.~18, pp.~3626--3649, 2013.

\bibitem{qin2021acoustic}
X.~Qin, X.~Wei, L.~Li, H.~Wang, Z.~Jiang, and D.~Sun, ``Acoustic valves in microfluidic channels for droplet manipulation,'' {\em Lab on a Chip}, vol.~21, no.~16, pp.~3165--3173, 2021.

\bibitem{zhao2025topological}
S.~Zhao, Z.~Tian, C.~Shen, S.~Yang, J.~Xia, T.~Li, Z.~Xie, P.~Zhang, L.~P. Lee, S.~A. Cummer, {\em et~al.}, ``Topological acoustofluidics,'' {\em Nature Materials}, pp.~1--9, 2025.

\bibitem{li2015acoustic}
P.~Li, Z.~Mao, Z.~Peng, L.~Zhou, Y.~Chen, P.-H. Huang, C.~I. Truica, J.~J. Drabick, W.~S. El-Deiry, M.~Dao, {\em et~al.}, ``Acoustic separation of circulating tumor cells,'' {\em Proceedings of the National Academy of Sciences}, vol.~112, no.~16, pp.~4970--4975, 2015.

\bibitem{pirkkalainen2013hybrid}
J.-M. Pirkkalainen, S.~Cho, J.~Li, G.~Paraoanu, P.~Hakonen, and M.~Sillanp{\"a}{\"a}, ``Hybrid circuit cavity quantum electrodynamics with a micromechanical resonator,'' {\em Nature}, vol.~494, no.~7436, pp.~211--215, 2013.

\bibitem{gustafsson2014propagating}
M.~V. Gustafsson, T.~Aref, A.~F. Kockum, M.~K. Ekstr{\"o}m, G.~Johansson, and P.~Delsing, ``Propagating phonons coupled to an artificial atom,'' {\em Science}, vol.~346, no.~6206, pp.~207--211, 2014.

\bibitem{whiteley2019spin}
S.~J. Whiteley, G.~Wolfowicz, C.~P. Anderson, A.~Bourassa, H.~Ma, M.~Ye, G.~Koolstra, K.~J. Satzinger, M.~V. Holt, F.~J. Heremans, {\em et~al.}, ``Spin--phonon interactions in silicon carbide addressed by gaussian acoustics,'' {\em Nature Physics}, vol.~15, no.~5, pp.~490--495, 2019.

\bibitem{maity2020coherent}
S.~Maity, L.~Shao, S.~Bogdanovi{\'c}, S.~Meesala, Y.-I. Sohn, N.~Sinclair, B.~Pingault, M.~Chalupnik, C.~Chia, L.~Zheng, {\em et~al.}, ``Coherent acoustic control of a single silicon vacancy spin in diamond,'' {\em Nature communications}, vol.~11, no.~1, p.~193, 2020.

\bibitem{manenti2017circuit}
R.~Manenti, A.~F. Kockum, A.~Patterson, T.~Behrle, J.~Rahamim, G.~Tancredi, F.~Nori, and P.~J. Leek, ``Circuit quantum acoustodynamics with surface acoustic waves,'' {\em Nature communications}, vol.~8, no.~1, p.~975, 2017.

\bibitem{schutz2017universal}
M.~J. Sch{\"u}tz and M.~J. Sch{\"u}tz, ``Universal quantum transducers based on surface acoustic waves,'' {\em Quantum dots for quantum information processing: controlling and exploiting the quantum dot environment}, pp.~143--196, 2017.

\bibitem{decrescent2022large}
R.~A. DeCrescent, Z.~Wang, P.~Imany, R.~C. Boutelle, C.~A. McDonald, T.~Autry, J.~D. Teufel, S.~W. Nam, R.~P. Mirin, and K.~L. Silverman, ``Large single-phonon optomechanical coupling between quantum dots and tightly confined surface acoustic waves in the quantum regime,'' {\em Physical Review Applied}, vol.~18, no.~3, p.~034067, 2022.

\bibitem{zhou2024electrically}
Y.~Zhou, F.~Ruesink, M.~Pavlovich, R.~Behunin, H.~Cheng, S.~Gertler, A.~L. Starbuck, A.~J. Leenheer, A.~T. Pomerene, D.~C. Trotter, {\em et~al.}, ``Electrically interfaced brillouin-active waveguide for microwave photonic measurements,'' {\em Nature Communications}, vol.~15, no.~1, p.~6796, 2024.

\bibitem{arrangoiz2019resolving}
P.~Arrangoiz-Arriola, E.~A. Wollack, Z.~Wang, M.~Pechal, W.~Jiang, T.~P. McKenna, J.~D. Witmer, R.~Van~Laer, and A.~H. Safavi-Naeini, ``Resolving the energy levels of a nanomechanical oscillator,'' {\em Nature}, vol.~571, no.~7766, pp.~537--540, 2019.

\bibitem{sletten2019resolving}
L.~R. Sletten, B.~A. Moores, J.~J. Viennot, and K.~W. Lehnert, ``Resolving phonon fock states in a multimode cavity with a double-slit qubit,'' {\em Physical Review X}, vol.~9, no.~2, p.~021056, 2019.

\bibitem{qiao2023splitting}
H.~Qiao, {\'E}.~Dumur, G.~Andersson, H.~Yan, M.-H. Chou, J.~Grebel, C.~R. Conner, Y.~J. Joshi, J.~M. Miller, R.~G. Povey, {\em et~al.}, ``Splitting phonons: Building a platform for linear mechanical quantum computing,'' {\em Science}, vol.~380, no.~6649, pp.~1030--1033, 2023.

\bibitem{qiao2025acoustic}
H.~Qiao, Z.~Wang, G.~Andersson, A.~Anferov, C.~R. Conner, Y.~J. Joshi, S.~Li, J.~M. Miller, X.~Wu, H.~Yan, {\em et~al.}, ``Acoustic phonon phase gates with number-resolving phonon detection,'' {\em arXiv preprint arXiv:2503.03898}, 2025.

\bibitem{shin2015control}
H.~Shin, J.~A. Cox, R.~Jarecki, A.~Starbuck, Z.~Wang, and P.~T. Rakich, ``Control of coherent information via on-chip photonic--phononic emitter--receivers,'' {\em Nature communications}, vol.~6, no.~1, p.~6427, 2015.

\bibitem{zhou2024nonreciprocal}
Y.~Zhou, F.~Ruesink, S.~Gertler, H.~Cheng, M.~Pavlovich, E.~Kittlaus, A.~L. Starbuck, A.~J. Leenheer, A.~T. Pomerene, D.~C. Trotter, {\em et~al.}, ``Nonreciprocal dissipation engineering via strong coupling with a continuum of modes,'' {\em Physical Review X}, vol.~14, no.~2, p.~021002, 2024.

\bibitem{freedman2025gigahertz}
J.~M. Freedman, M.~J. Storey, D.~Dominguez, A.~Leenheer, S.~Magri, N.~T. Otterstrom, and M.~Eichenfield, ``Gigahertz-frequency, acousto-optic phase modulation of visible light in a cmos-fabricated photonic circuit,'' {\em arXiv preprint arXiv:2502.08012}, 2025.

\bibitem{shao2019microwave}
L.~Shao, M.~Yu, S.~Maity, N.~Sinclair, L.~Zheng, C.~Chia, A.~Shams-Ansari, C.~Wang, M.~Zhang, K.~Lai, {\em et~al.}, ``Microwave-to-optical conversion using lithium niobate thin-film acoustic resonators,'' {\em Optica}, vol.~6, no.~12, pp.~1498--1505, 2019.

\bibitem{hassanien2021efficient}
A.~E. Hassanien, S.~Link, Y.~Yang, E.~Chow, L.~L. Goddard, and S.~Gong, ``Efficient and wideband acousto-optic modulation on thin-film lithium niobate for microwave-to-photonic conversion,'' {\em Photonics Research}, vol.~9, no.~7, pp.~1182--1190, 2021.

\bibitem{kittlaus2021electrically}
E.~A. Kittlaus, W.~M. Jones, P.~T. Rakich, N.~T. Otterstrom, R.~E. Muller, and M.~Rais-Zadeh, ``Electrically driven acousto-optics and broadband non-reciprocity in silicon photonics,'' {\em Nature Photonics}, vol.~15, no.~1, pp.~43--52, 2021.

\bibitem{li2023frequency}
B.~Li, Q.~Lin, and M.~Li, ``Frequency--angular resolving lidar using chip-scale acousto-optic beam steering,'' {\em Nature}, vol.~620, no.~7973, pp.~316--322, 2023.

\bibitem{lin2024optical}
Q.~Lin, S.~Fang, Y.~Yu, Z.~Xi, L.~Shao, B.~Li, and M.~Li, ``Optical multi-beam steering and communication using integrated acousto-optics arrays,'' {\em arXiv preprint arXiv:2409.16511}, 2024.

\bibitem{zhao2022enabling}
H.~Zhao, B.~Li, H.~Li, and M.~Li, ``Enabling scalable optical computing in synthetic frequency dimension using integrated cavity acousto-optics,'' {\em Nature Communications}, vol.~13, no.~1, p.~5426, 2022.

\bibitem{neuman2021phononic}
T.~Neuman, M.~Eichenfield, M.~E. Trusheim, L.~Hackett, P.~Narang, and D.~Englund, ``A phononic interface between a superconducting quantum processor and quantum networked spin memories,'' {\em npj Quantum Information}, vol.~7, no.~1, p.~121, 2021.

\bibitem{lu2012novel}
X.~Lu, J.~Ma, X.~L. Zhu, C.~M. Lee, C.~P. Yue, and K.~M. Lau, ``A novel gan-based monolithic saw/hemt oscillator on silicon,'' in {\em 2012 IEEE International Ultrasonics Symposium}, pp.~2206--2209, IEEE, 2012.

\bibitem{bahr201516}
B.~W. Bahr, L.~C. Popa, and D.~Weinstein, ``16.8 1ghz gan-mmic monolithically integrated mems-based oscillators,'' in {\em 2015 IEEE International Solid-State Circuits Conference-(ISSCC) Digest of Technical Papers}, pp.~1--3, IEEE, 2015.

\bibitem{xi2025low}
Z.~Xi, J.~G. Thomas, J.~Ji, D.~Wang, Z.~Cen, I.~I. Kravchenko, B.~R. Srijanto, Y.~Yao, Y.~Zhu, and L.~Shao, ``Low-phase-noise surface-acoustic-wave oscillator using an edge mode of a phononic band gap,'' {\em Physical Review Applied}, vol.~23, no.~2, p.~024054, 2025.

\bibitem{vahala2009phonon}
K.~Vahala, M.~Herrmann, S.~Kn{\"u}nz, V.~Batteiger, G.~Saathoff, T.~H{\"a}nsch, and T.~Udem, ``A phonon laser,'' {\em Nature Physics}, vol.~5, no.~9, pp.~682--686, 2009.

\bibitem{grudinin2010phonon}
I.~S. Grudinin, H.~Lee, O.~Painter, and K.~J. Vahala, ``Phonon laser action in a tunable two-level system,'' {\em Physical review letters}, vol.~104, no.~8, p.~083901, 2010.

\bibitem{zhang2018phonon}
J.~Zhang, B.~Peng, {\c{S}}.~K. {\"O}zdemir, K.~Pichler, D.~O. Krimer, G.~Zhao, F.~Nori, Y.-x. Liu, S.~Rotter, and L.~Yang, ``A phonon laser operating at an exceptional point,'' {\em Nature Photonics}, vol.~12, no.~8, pp.~479--484, 2018.

\bibitem{pettit2019optical}
R.~M. Pettit, W.~Ge, P.~Kumar, D.~R. Luntz-Martin, J.~T. Schultz, L.~P. Neukirch, M.~Bhattacharya, and A.~N. Vamivakas, ``An optical tweezer phonon laser,'' {\em Nature Photonics}, vol.~13, no.~6, pp.~402--405, 2019.

\bibitem{iyer2024coherent}
A.~Iyer, Y.~P. Kandel, W.~Xu, J.~M. Nichol, and W.~H. Renninger, ``Coherent optical coupling to surface acoustic wave devices,'' {\em Nature Communications}, vol.~15, no.~1, p.~3993, 2024.

\bibitem{okada1964continuous}
J.~Okada and H.~Matino, ``Continuous oscillations of acoustoelectric current in cds,'' {\em Japanese Journal of Applied Physics}, vol.~3, no.~11, p.~698, 1964.

\bibitem{maines1968gigahertz}
J.~Maines, F.~Marshall, E.~Paige, and R.~Stuart, ``Gigahertz acousto-electric oscillations in zinc oxide,'' {\em Physics Letters A}, vol.~26, no.~8, pp.~388--389, 1968.

\bibitem{maines1970current}
J.~Maines and E.~Paige, ``Current-spiking and self-locking of modes of the acousto-electric oscillator,'' {\em Solid State Communications}, vol.~8, no.~6, pp.~421--425, 1970.

\bibitem{gokhale2014phonon}
V.~J. Gokhale and M.~Rais-Zadeh, ``Phonon-electron interactions in piezoelectric semiconductor bulk acoustic wave resonators,'' {\em Scientific reports}, vol.~4, no.~1, p.~5617, 2014.

\bibitem{mansoorzare2019acoustoelectric}
H.~Mansoorzare and R.~Abdolvand, ``Acoustoelectric amplification in lateral-extensional composite piezo-silicon resonant cavities,'' in {\em 2019 Joint Conference of the IEEE International Frequency Control Symposium and European Frequency and Time Forum (EFTF/IFC)}, pp.~1--3, IEEE, 2019.

\bibitem{hackett2023non}
L.~Hackett, M.~Miller, S.~Weatherred, S.~Arterburn, M.~J. Storey, G.~Peake, D.~Dominguez, P.~S. Finnegan, T.~A. Friedmann, and M.~Eichenfield, ``Non-reciprocal acoustoelectric microwave amplifiers with net gain and low noise in continuous operation,'' {\em Nature Electronics}, vol.~6, no.~1, pp.~76--85, 2023.

\bibitem{hackett2021towards}
L.~Hackett, M.~Miller, F.~Brimigion, D.~Dominguez, G.~Peake, A.~Tauke-Pedretti, S.~Arterburn, T.~A. Friedmann, and M.~Eichenfield, ``Towards single-chip radiofrequency signal processing via acoustoelectric electron--phonon interactions,'' {\em Nature communications}, vol.~12, no.~1, p.~2769, 2021.

\bibitem{hackett2024giant}
L.~Hackett, M.~Koppa, B.~Smith, M.~Miller, S.~Santillan, S.~Weatherred, S.~Arterburn, T.~A. Friedmann, N.~Otterstrom, and M.~Eichenfield, ``Giant electron-mediated phononic nonlinearity in semiconductor--piezoelectric heterostructures,'' {\em Nature Materials}, vol.~23, no.~10, pp.~1386--1393, 2024.

\bibitem{kino1971normal}
G.~Kino and T.~Reeder, ``A normal mode theory for the rayleigh wave amplifier,'' {\em IEEE Transactions on Electron Devices}, vol.~18, no.~10, pp.~909--920, 1971.

\bibitem{pippard1963acoustic}
A.~Pippard, ``Acoustic amplification in semiconductors and metals,'' {\em Philosophical Magazine}, vol.~8, no.~85, pp.~161--165, 1963.

\bibitem{coldren1972monolithic}
L.~A. Coldren, {\em Monolithic acoustic surface wave amplifiers}.
\newblock Stanford University, 1972.

\bibitem{rhea1990oscillator}
R.~W. Rhea, {\em Oscillator design \& computer simulation}.
\newblock Prentice-Hall, Inc., 1990.

\bibitem{danicki1993reversing}
E.~Danicki, ``Reversing multistrip coupler,'' {\em Ultrasonics}, vol.~31, no.~6, pp.~421--424, 1993.

\bibitem{bajak1981attenuation}
I.~Bajak, A.~McNab, J.~Richter, and C.~Wilkinson, ``Attenuation of acoustic waves in lithium niobate,'' {\em The Journal of the Acoustical Society of America}, vol.~69, no.~3, pp.~689--695, 1981.

\bibitem{xie2024high}
J.~Xie, M.~Shen, and H.~X. Tang, ``High acoustic velocity x-cut lithium niobate sub-terahertz electromechanics,'' {\em Applied Physics Letters}, vol.~124, no.~7, 2024.

\bibitem{shao2019phononic}
L.~Shao, S.~Maity, L.~Zheng, L.~Wu, A.~Shams-Ansari, Y.-I. Sohn, E.~Puma, M.~Gadalla, M.~Zhang, C.~Wang, {\em et~al.}, ``Phononic band structure engineering for high-q gigahertz surface acoustic wave resonators on lithium niobate,'' {\em Physical Review Applied}, vol.~12, no.~1, p.~014022, 2019.

\bibitem{chatterjee2024ab}
E.~Chatterjee, A.~Wendt, D.~Soh, and M.~Eichenfield, ``Ab initio calculations of nonlinear susceptibility and multiphonon mixing processes in a 2deg-piezoelectric heterostructure,'' {\em Physical Review Research}, vol.~6, no.~2, p.~023288, 2024.

\bibitem{keysight_phase_noise}
{Keysight Technologies}, ``Measuring phase noise with a real-time sampling oscilloscope,'' 2025.
\newblock Accessed: 2025-04-17.

\end{thebibliography}


\begin{thebibliography}{10}

\bibitem{morgan2010surface}
D.~Morgan, {\em Surface acoustic wave filters: With applications to electronic communications and signal processing}.
\newblock Academic Press, 2010.

\bibitem{hackett2023non}
L.~Hackett, M.~Miller, S.~Weatherred, S.~Arterburn, M.~J. Storey, G.~Peake, D.~Dominguez, P.~S. Finnegan, T.~A. Friedmann, and M.~Eichenfield, ``Non-reciprocal acoustoelectric microwave amplifiers with net gain and low noise in continuous operation,'' {\em Nature Electronics}, vol.~6, no.~1, pp.~76--85, 2023.

\bibitem{shao2019phononic}
L.~Shao, S.~Maity, L.~Zheng, L.~Wu, A.~Shams-Ansari, Y.-I. Sohn, E.~Puma, M.~Gadalla, M.~Zhang, C.~Wang, {\em et~al.}, ``Phononic band structure engineering for high-q gigahertz surface acoustic wave resonators on lithium niobate,'' {\em Physical Review Applied}, vol.~12, no.~1, p.~014022, 2019.

\bibitem{ward1990temperature}
R.~B. Ward, ``Temperature coefficients of saw delay and velocity for y-cut and rotated linbo/sub 3,'' {\em IEEE transactions on ultrasonics, ferroelectrics, and frequency control}, vol.~37, no.~5, pp.~481--483, 1990.

\bibitem{cho1987nonlinear}
Y.~Cho and K.~Yamanouchi, ``Nonlinear, elastic, piezoelectric, electrostrictive, and dielectric constants of lithium niobate,'' {\em Journal of applied physics}, vol.~61, no.~3, pp.~875--887, 1987.

\bibitem{schawlow1958infrared}
A.~L. Schawlow and C.~H. Townes, ``Infrared and optical masers,'' {\em Physical review}, vol.~112, no.~6, p.~1940, 1958.

\bibitem{henry1982theory}
C.~Henry, ``Theory of the linewidth of semiconductor lasers,'' {\em IEEE Journal of Quantum Electronics}, vol.~18, no.~2, pp.~259--264, 1982.

\bibitem{gokhale2014phonon}
V.~J. Gokhale and M.~Rais-Zadeh, ``Phonon-electron interactions in piezoelectric semiconductor bulk acoustic wave resonators,'' {\em Scientific reports}, vol.~4, no.~1, p.~5617, 2014.

\bibitem{rhea1990oscillator}
R.~W. Rhea, {\em Oscillator design \& computer simulation}.
\newblock Prentice-Hall, Inc., 1990.

\bibitem{hackett2021towards}
L.~Hackett, M.~Miller, F.~Brimigion, D.~Dominguez, G.~Peake, A.~Tauke-Pedretti, S.~Arterburn, T.~A. Friedmann, and M.~Eichenfield, ``Towards single-chip radiofrequency signal processing via acoustoelectric electron--phonon interactions,'' {\em Nature communications}, vol.~12, no.~1, p.~2769, 2021.

\bibitem{hackett2024giant}
L.~Hackett, M.~Koppa, B.~Smith, M.~Miller, S.~Santillan, S.~Weatherred, S.~Arterburn, T.~A. Friedmann, N.~Otterstrom, and M.~Eichenfield, ``Giant electron-mediated phononic nonlinearity in semiconductor--piezoelectric heterostructures,'' {\em Nature Materials}, vol.~23, no.~10, pp.~1386--1393, 2024.

\bibitem{ismail2016fabry}
N.~Ismail, C.~C. Kores, D.~Geskus, and M.~Pollnau, ``Fabry-p{\'e}rot resonator: spectral line shapes, generic and related airy distributions, linewidths, finesses, and performance at low or frequency-dependent reflectivity,'' {\em Optics express}, vol.~24, no.~15, pp.~16366--16389, 2016.

\end{thebibliography}
%\newpage
\section{Methods}
\subsection{Device fabrication}
A lattice-matched epitaxial semiconductor layer is first grown on a two inch Indium Phosphide (InP) wafer using metal-organic chemical vapor deposition, consisting of a 500 nm non-intentionally doped InP buffer, a 3000 nm non-intentionally doped In$_{0.53}$Ga$_{0.47}$As etch stop, a 100 nm InP etch stop doped with silicon at $N_d = 1\times10^{18}$cm$^{-3}$, a two-layer epitaxial contact (100-nm-thick In$_{0.53}$Ga$_{0.47}$As and 30-nm-thick InP contact layers doped with silicon at $N_d=2\times10^{19}$cm$^{-3}$ and $N_d=1\times10^{18}$cm$^{-3}$, respectively), an In$_{0.53}$Ga$_{0.47}$As amplifier layer with a target silicon doping level of $N_d=5\times10^{15}$cm$^{-3}$ and a target thickness of 50 nm, and a 5-nm-thick non-intentionally doped InP adhesion layer. The InP wafer is then bonded to a four-inch 5-\si{\micro\meter} LN on silicon wafer, initiated manually followed by annealing in vacuum at \ang{100} C. After bonding, the InP substrate and buffer layer are etched away using hydrochloric acid. Then, the In$_{0.53}$Ga$_{0.47}$As etch-stop layer is etched using a 1:1:10 solution of sulfuric acid, hydrogen peroxide and water. The InP etch-stop layer is then removed in a 1:3 solution of hydrochloric acid and phosphoric acid. The In$_{0.53}$Ga$_{0.47}$As contact layer is then patterned and subsequently etched in a 1:1:10 solution of sulfuric acid, hydrogen peroxide and water, immediately followed by etching the subsequent InP etch-stop layer in a 1:3 mixture of hydrochloric acid/phosphoric acid. The In$_{0.53}$Ga$_{0.47}$As amplifier layer is then patterned and etched in a 1:1:10 solution of sulfuric acid, hydrogen peroxide and water, landing on the LN. The IDTs are formed using a metal liftoff process, using 10 nm titanium and 100 nm aluminum, with a thin gold capping layer to prevent oxidation. A second liftoff process of 20 nm titanium, 500 nm gold, 500 nm silver, and 100 nm gold is used to form the contact to the In$_{0.53}$Ga$_{0.47}$As contact layer.

\subsection{Data acquisition}
All data was taken on a custom RF-DC probe station. Scattering (S) parameters for the passive resonator and resonant amplifier data were taken using a Keysight P9374A Vector Network Analyzer (VNA), making contact to the IDTs that probe the devices using GSG probes from FormFactor. A two-port short-load-open-through calibration was performed using a calibration substrate prior to data collection. Passive resonator data was taken using a VNA frequency sweep. Resonant amplifier data was taken by applying a continuous drift field using a DC power supply (Keithely 2450 Source Measurement Unit), and measuring the S-parameters as a function of frequency using the VNA. For each measurement, a voltage was set on the power supply, and a frequency sweep was subsequently performed by the VNA. This sweep takes on the order of seconds to perform, significantly longer than any transient response we expect to see in our system \cite{hackett2023non}.

All oscillator measurements other than the broad spectrum (Fig. 3g) were taken using a Pico Technology Picoscope 6428E-D oscilloscope, operating at a sampling rate of 5 GS/s to resolve time domain signals below 2.5 GHz. The power out vs. bias power data (Fig. 3b) was taken by setting a bias voltage, waiting a second to allow the current to settle, then capture 2 ms of data, or $10^7$ data points. After saving the data from the scope and recording the bias voltage and current, the next bias was set and the process was repeated. After data acquisition, the power spectral density (PSD) of each file was calculated using Welch's method. A trapezoidal integration was subsequently applied in the band from 993 to 994 MHz to capture the power contained in the central peak. This power was then scaled by the correction factor, explained in depth in supplemental note B.

The linewidth narrowing plot was taken by applying a linear DC bias ramp from 33 V to 36 V over the course of 200 ms. This signal was captured by the oscilloscope, after which the data was post processed by splitting the dataset into 5 ms chunks. Each chunk had a Hann window applied, and an FFT subsequently taken. The applied DC bias was back calculated, resulting in a range of voltages within that specific time frame. The values displayed are the average of these voltages, rounded to the nearest .1 V. The voltages changed by approximately 0.079 V across each frame, about $10\%$ of the difference of voltage between displayed frames. Several of these post processed frames were then selected to be overlaid to create the final figure. No correction factor was applied to this data. 

The linewidth data was taken by applying a continuous DC bias and capturing data with the oscilloscope for 200 ms. This data was then parsed into a 20 ms long subsection where the oscillation frequency was observed to be the most stable, 78 ms to 98 ms. A Hann window was then applied to the data, after which was zero padded to achieve a 5 Hz resolution. A Gaussian fit was then calculated, which we found to be better than a Lorentzian or Voigt fit. The power was calculated by using Welch's method on the same data subset and integrating over a span of 1 kHz around the center frequency. Afterwards, the correction factor was applied to get the final result of -6.1 dBm.

The phase noise was calculated from the same set of data as the linewidth measurement. After applying a Hann window to the entire 200 ms, the phase noise was calculated using the IQ demodulation technique \cite{keysight_phase_noise}. The data was digitally mixed with a signal of the same frequency as the center frequency of the entire waveform. After which, the I and Q signals were loss pass filtered with a butterworth filter at 500 MHz. The phase was calculated using the arctan of these signals. After removing a linear trend, Welch's method was used to calculate the PSD, which was converted to units of dBc/Hz and plotted. The fit was found by calculating the power, effective quality factor, using a previous result for the noise factor, and fitting the flicker corner. 

The frequency tuning measurement was captured in the same way as the linewidth narrowing plot, but sweeping the DC bias from 33 V to 40 V. The data was then split into 5 ms chunks, each of which had a Hann window applied. The maximum peak of each window was found and plotted. These were then linearly fit.

The spectrum data was taken using a Keysight N9000B CXA Signal Analyzer (ESA), using a 100 Hz resolution bandwidth. A continuous DC bias was set, and a frequency sweep was performed using the ESA. The data was post processed to account of insertion loss and aperture mismatch, details in supplemental note C.

\section{Acknowledgments}
This material is based on research sponsored in part by the Defense Advanced Research Projects Agency (DARPA) through a Young Faculty Award (YFA) under grant D23AP00174-00. The views and conclusions contained herein are those of the authors and should not be interpreted as necessarily representing the oficial policies or endorsements, either expressed or implied, by DARPA, the Department of the Interior, or the US Government. This work is supported by the Laboratory Directed Research and Development program at Sandia National Laboratories, a multimission laboratory managed and operated by National Technology and Engineering Solutions of Sandia LLC, a wholly owned subsidiary of Honeywell International Inc. for the US Department of Energy’s National Nuclear Security Administration under contract DE-NA0003525. This work was performed, in part, at the Center for Integrated Nanotechnologies, an Office of Science User Facility, operated for the US Department of Energy Office of Science. This paper describes objective technical results and analysis. Any subjective views or opinions that might be expressed in the paper do not necessarily represent the views of the US Department of Energy or the US Government.

\end{document}

\preprint{APS/123-QED}

\title{An Electrically Injected and Solid State Surface Acoustic Wave Phonon Laser}% Force line breaks with \\
%\affiliation{College of Optical Sciences, SNL
%}%


\date{\today}% It is always \today, today,

\section{Supplemental for "An Electrically Injected and Solid State Surface Acoustic Wave Phonon Laser"}
\subsection{Stimulated emission and absorption}
While the 1-D example described in the main text is sufficient to illustrate the basic principle of stimulated emission and absorption, higher dimensional analysis is required to explain how the rates of these process change in response to an applied bias. An example gain curve is shown in supplemental figure 1a, using different parameters than the real device to exaggerate its features. For ease of understanding and clarity of illustration, we will limit this discussion to two dimensions, although the fundamental logic carries to three dimensions.
As in the main text, consider the interaction of a phonon with k-vector $\textbf{q} = +q\hat{x}$ with an electron gas. For simplicity, we will initially assume zero temperature, ignore the lifetime-limited energy broadening of the electronic states, and assume the electronic mean free path is sufficiently long enough such that $ql \gg 1$, where $l$ is the electron mean free path. The interaction between the phonon and the electron gas must conserve energy and momentum. In one dimension, there is exactly one pair of states that can satisfy this conservation requirement. If the lower energy state is occupied and the higher state is unoccupied, the phonon may be absorbed. In the case of an occupied higher state and unoccupied lower state, the phonon may cause stimulated emission. Should both states be occupied or unoccupied, the phonon is unable to interact with the gas.

In two dimensions, there are many pairs of states that can satisfy this requirement, as multiple states can have the same $k_x$ so long as they differ in $k_y$. A comparison of the 1-D and 2-D cases is shown in supplemental figure 1b. In one dimension, the electronic states form a parabola, while in two dimensions, the states form a paraboloid. For simplicity, we will assume equal dimensions in the X and Y directions, making it a circular paraboloid. By projecting the energy axis of the dispersion diagram into the page, we may draw a circle to represent the occupied electronic states, with the white space outside of this unoccupied states. In the two dimensional diagrams, the set of initial states is shown by the dark blue line, the final set of states represented with a light blue line, and the gray arrow indicates the transition from one to the other. The presence of a DC bias $E = E_x \hat{x}$ shifts the average momentum of the electronic states to $\textbf{k} = m_e \mu E_x / \hbar \hat{x} = k_x \hat{x}$, resulting in a shift of the center of the occupation circle to $k_x \hat{x}$.

The rate of absorption/emission depends on the number potential transition pairs. Supplemental figure 1c depicts the effect of increasing DC bias, steadily moving the center to the right. At zero bias, the circle of occupation is centered at the origin. In this illustration, all of the available initial states have unoccupied final states. As the bias is increased, the final states become increasingly occupied, represented by the dotted lines inside the occupation circle. These occupied states are no longer free as final states, and thus the total population of available transition pairs goes down, lowering the rate of absorption. Eventually, all the potential initial states have occupied final states, at which point there no absorption or emission. Continuing to increase the bias past this will free up low energy states and populate more high energy states, creating a population inversion and allowing for stimulated emission.

The behavior of acoustoelectric loss for counter-propagating phonons may also be intuitively explained using this model, shown in supplemental figure 1d. Phonons propagating with k-vector $-q$ will excite low-energy electrons on the right of the distribution to unoccupied, high energy states on the left of the distribution, the mirror of forward propagating phonons. As the bias increases, the number of available transition pairs also increases, until the circle of occupation is centered at the $k_x$ of the lower energy state of the transition. After this point, increasing the bias decreases the number available states, decreasing the rate of absorption.

The condition for the gain to overcome the loss is illustrated in supplemental figure 1e. Here, the dark purple and light purple lines represent initial and final states of the backward propagating phonons, while the dark and light blue states represent the initial and final states of the forward propagating phonons. The bias must be increased past the point of peak loss, after which the loss will begin to decrease while the gain will continue to rise. The final figure shows that the states available for emission in the forward direction outnumber the states that can absorb backward phonons, seen from the longer dark blue line than the dark purple line.

The effect of carrier density may also be intuitively visualized. The radius of the circle of occupation increases as the carrier density increases. The consequences of this can be complicated and depend on the specific interaction. Supplemental figure 2a depicts the effect of carrier concentration on stimulated emission at a particular bias. We can see that increasing the number of carriers will increase the number of available high energy states, until a certain point where the addition of carriers will begin to occupy the low energy states, reducing the emission rate. Thus, to reach peak gain, the bias must be increased, but the peak amplification will be larger. 

If the electron gas has non-zero temperature, the occupation of electronic states becomes probabilistic. In this depiction, this effectively blurs out the edges of the circle, shown in supplemental figure 2b. This also now permits absorption and emission to occur simultaneously. As is the case of lasers, the transition probability from a fully unoccupied initial to a fully unoccupied final state is the same for both emission and absorption, so the net effect depends on the occupation numbers of the higher or lower energy states. Additionally, this increase in states available for stimulated emission also gives rise to an increased rate of spontaneous emission, increasing the noise.

\begin{figure*}[htbp]
  \centering
  \includegraphics[width=\linewidth]{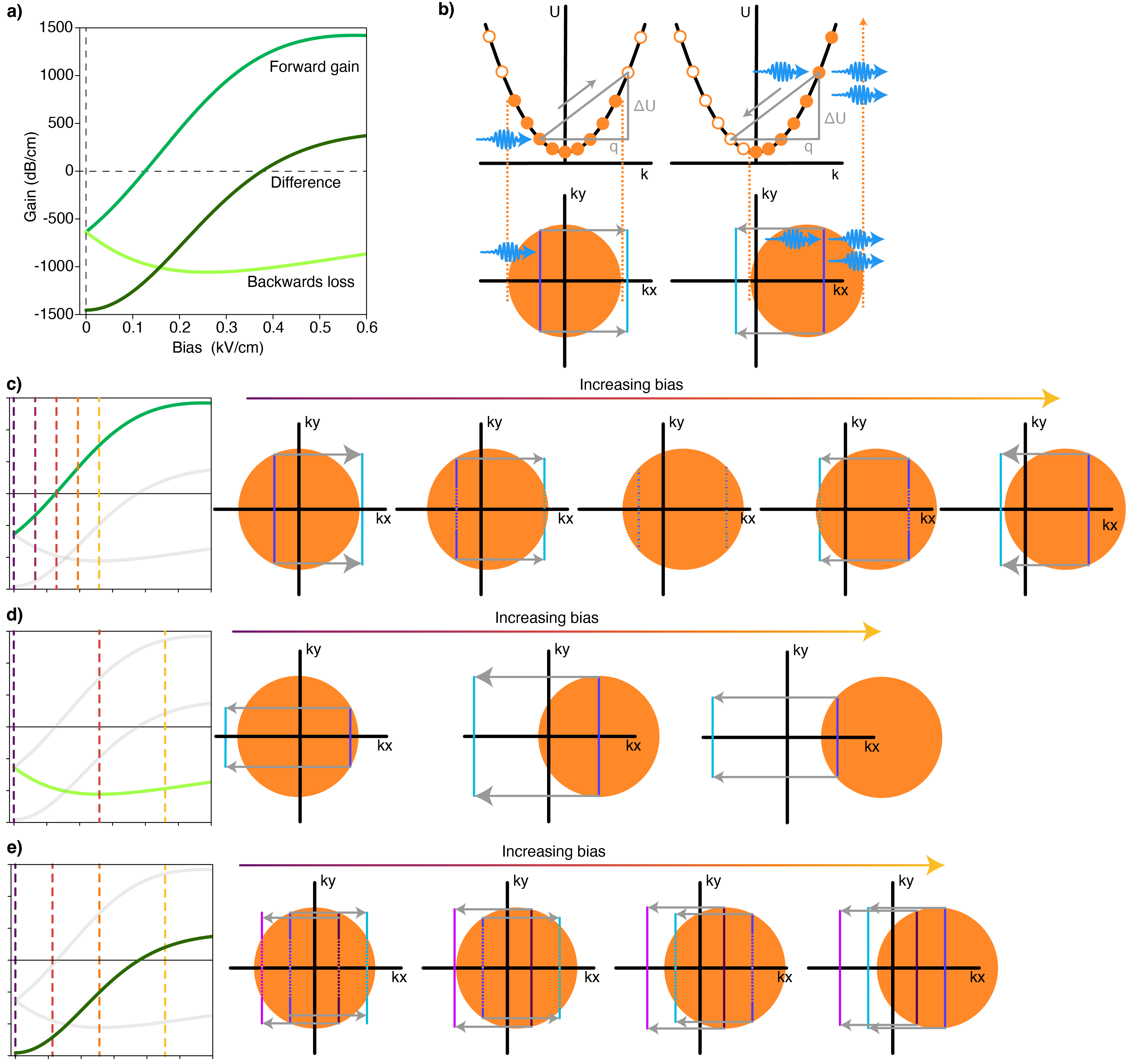}
  \caption{
  Stimulated emission and absorption diagrams. Circles represent the set of occupied states, dark blue lines represent initial states, and light blue lines represent final states. a) Example classical gain curve for reference. b) Extrapolation from 1D to 2D. c) Increasing DC bias from left to right, resulting in a change of the population of initial and final states, and ultimately transitioning from stimulated absorption to emission. Dotted lines along gain curve correspond with each diagram d) Increasing bias increases loss experienced by counter-propagating phonons, until the number of initial states reduces. e) Gain and loss at the same bias. Dark red lines represent initial states and pink lines represent final states for the absorption of backward propagating phonons. At high enough bias, we see that the number of states available for gain of forward phonons is greater than the number of states available for absorption of backward phonons, allowing for round-trip gain in a Fabry-Perot resonator.
  }
  \label{fig:supp: stim emission}
\end{figure*}

\begin{figure*}[htbp]
  \centering
  \includegraphics[width=\linewidth]{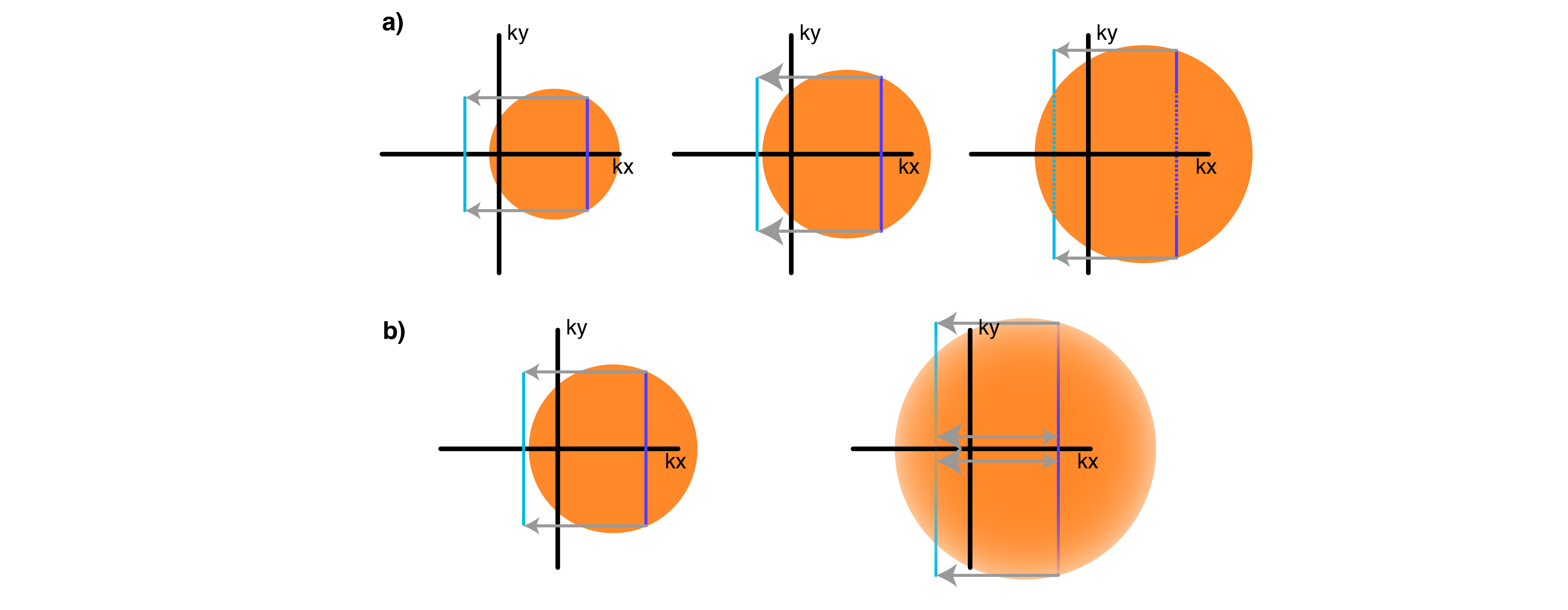}
  \caption{
  a) Increasing the carrier density increases the radius of the circle of occupation. b) Right diagram shows the effects of finite temperature blurring of the edge of the circle. 
  }
  \label{fig:supp: stim emission other}
\end{figure*}

In a real system, both the k-vectors and energy levels will be broadened. The finite confinement of the electrons localize the electrons, leading to broadened k-vectors. Additionally, collisions in the semiconductor responsible for limiting the velocity of the electrons to the drift velocity ultimately causes broadening of the energy levels of the electronic states. A finite temperature will contribute to both k-vector and energy broadening, following fermi-dirac statistics. Other processes may affect the system depending on the specific material, frequency, and temperature. These broadening mechanisms will relax the energy and momentum conservation criteria, making off-resonant transitions possible.

\subsection{Acoustoelectric modeling}
A schematic of the amplifier heterostructure is shown in supplemental figure 3a. A thin dielectric layer of thickness $d$ and permittivity $\epsilon_d$ separates the piezoelectric layer from the amplifying semiconductor layer. Following the work of Kino and Reeder and considering the modifications made by Coldren, we use the following as the analytical expressions for the modification of the acoustic propagation constant

\begin{align}
    g_{ae} = \frac{1}{2} k^2_{\text{eff}}\beta_a\frac{(v_d/v_a - 1)(\omega_c/\omega)}{(v_d/v_a - 1)^2 + (\frac{R\omega_c}{\omega} + H)^2}
    \label{eq: ae gain}
\end{align}
\begin{align}
    \beta_{ae} = \frac{1}{2} k^2_{\text{eff}}\beta_a \frac{(R\omega_c/\omega + H)(\omega_c/\omega)}{(v_d/v_a - 1)^2 + (\frac{R\omega_c}{\omega} + H)^2}
    \label{eq: ae prop}
\end{align}
where $v_d$ is the electron drift velocity, $v_a$ is the acoustic velocity, and $\omega$ is the acoustic frequency. The dielectric relaxation frequency $\omega_c$ is defined to be
\begin{align}
    \omega_c = \mu e N/\epsilon_s,
\end{align}
where $\mu$ is the semiconductor mobility, $e$ is the electron charge, and $N$ is the charge carrier density. Above this frequency, the wavelength is sufficiently short enough such that diffusion will smooth out the carrier density fluctuations, leading to a decrease in gain. The effective electromechanical coupling coefficient $k^2_{\text{eff}}$ is defined as
\begin{align}
    k^2_{\text{eff}} &=  \epsilon_s \omega  wZ_a' \tanh(\beta_a d), \\
    wZ_a' &= wZ_{a0}\big[1 - (\epsilon_d/\epsilon_h -1)M'(\beta h)]^2 e^{-2\beta_a h}, \\
    wZ_{a0} &= \frac{k^2}{\omega(\epsilon_h+\epsilon_p)},
\end{align}
where $k^2 = 2 |\frac{\Delta v}{v}|$ is the typical electromechanical coupling coefficient. The presence of the dielectric gap between the piezoelectric and semiconductor layers modifies the coupling, accounted for with the modified space-charge potential factor $M'$, defined as
\begin{align}
    M'(\beta h) &= \frac{M(\beta h)}{1 + (\epsilon_d/\epsilon_h - 1)M(\beta h)},
\end{align}    
where
\begin{align}
    M(\beta h) &= \frac{\epsilon_h + \epsilon_p \tanh \beta_a h}{(\epsilon_h + \epsilon_p)(1+\tanh \beta_a h)}. \\
\end{align} 
Additionally, the space-charge reduction factor is written as
\begin{align}
    R &= (\epsilon_s / \epsilon_h) M'\tanh \beta_a d.
\end{align}
Finally, the diffusion term H is defined as
\begin{align}
    H &= \sqrt{\frac{\omega_c}{\omega_D}}\frac{\tanh \beta_a d}{\tanh d/\lambda_D},
\end{align}
where $\omega_D$ is the diffusion frequency,
\begin{align}
    \omega_D &= \frac{v_a^2}{(k_B T/q)\mu}.
\end{align}
In the regime where $\omega \ll \sqrt{\omega_c \omega_D}$, the Debye length $\lambda_d$ in the semiconductor may be written as
\begin{align}
    \lambda_d &= \frac{v_a}{\sqrt{\omega_c\omega_D}}
\end{align}

Optimization of the semiconductor parameters is a problem-specific task. In ring resonator topologies, the amount of round-trip loss is independent of the applied bias, so the semiconducting medium should be optimized to produce maximum gain, generally speaking. Increasing the mobility will decrease the bias required to achieve $v_0 = v_a$, the point at which the acoustoelectric effect becomes gainy, thus maximizing this will improve device performance. Carrier concentration has multiple effects that must be accounted for. Larger values will increase the maximum amount of gain, but simultaneously increase the bias fields required to obtain these peaks to impractically large values. Exceptionally small values ($<10^{15}$\si{\centi\meter^{-3}}) need much lower biases, but have the trade-off of a low gain slow and low peak gain. To overcome internal losses at the lowest possible bias, the gain slope should be maximized. For this particular heterostructure and mode at 1 GHz, the carrier concentration that gives a maximum gain slope is about $3\times10^{15}$\si{\centi\meter^{-3}}. These effects are shown in supplemental figure 3b, which depicts the bias field required for net round trip gain for a ring resonator topology, including propagation and mirror losses.

\begin{figure*}[htbp]
    \centering
    \includegraphics[width = \linewidth]{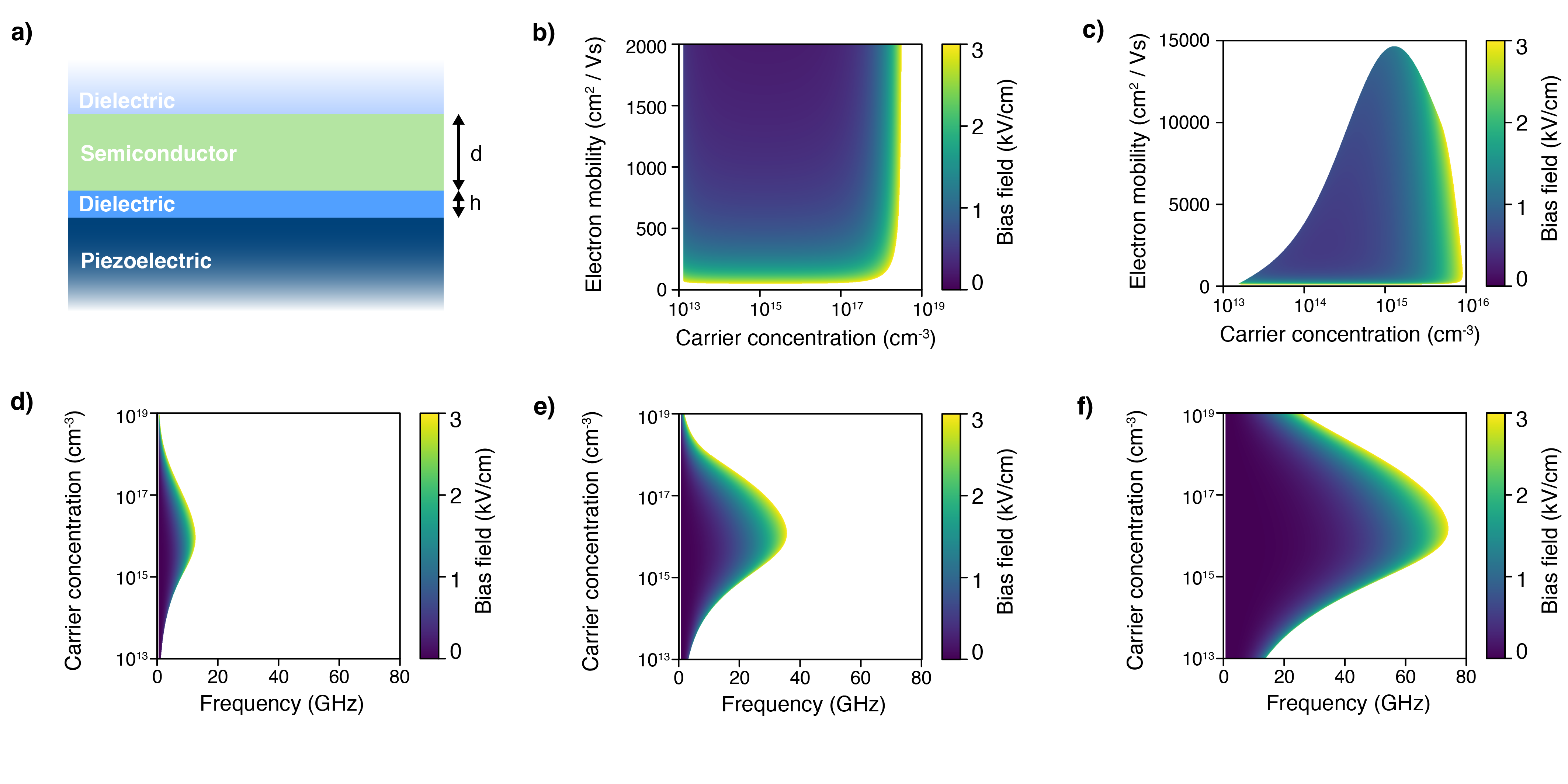}
    \caption{\textbf{Acoustoelectric modeling.} \textbf{a)} Diagram of heterostructure. \textbf{b)} Calculated threshold bias for a ring cavity assuming similar propagation and mirror losses. \textbf{c)} Threshold bias for a Fabry-Perot cavity. \textbf{d)} Threshold bias for a ring cavity at high frequencies for the Y-X LN q-SH0 mode. \textbf{e)} Threshold bias at high frequencies for the Y-Z LN rayleigh mode. \textbf{f)} Threshold bias at high frequencies for an X-cut LN plate mode.}
    \label{fig:supp: parameter fitting}
\end{figure*}

Increasing frequency has a similar effect as increasing the carrier concentration. Increasing frequency will in general increase the peak of the gain slope, but will also increase the bias required to obtain it, while decreasing it too low will decrease the gain slope and peak gain. Optimum frequency is highly dependent on the carrier concentration, and vice-versa. For example, at a carrier concentration of $1\times10^{17}$\si{\centi\meter^{-3}}, the maximum gain slope occurs at around 6.5 GHz, while a carrier concentration of $1\times10^{15}$\si{\centi\meter^{-3}} maximizes the gain slope of 500 MHz.

This optimization problem becomes more complicated for Fabry-Perot topologies, as increasing the gain will also increase the loss in the backward direction, and thus won't necessarily decrease the threshold voltage. In order for the gain to overcome the losses, the bias must overcome the negative peak of the loss, so in general decreasing the point at which the gain peaks is beneficial. Increasing mobility is not always beneficial, as while it will increase the gain, it will simultaneously increase the loss. These effects can be seen in supplemental figure 3c, which depicts the same system as 3b, but including the backwards acoustoelectric losses found in a Fabry-Perot topology.

As seen in the figure, it is usually beneficial to decrease the carrier concentration, as this will generally decrease the threshold bias. This will depend on system losses, however, as the maximum round-trip gain does decrease as the carrier concentration decreases, so care must be taken to ensure that the specific system isn't too lossy such that a lower carrier concentration prohibits net round-trip gain.

The choice of material platform is of great importance when scaling to high frequencies. Supplemental figures d-f depict the predicted threshold bias of a ring oscillator in three different material platforms: Y-X LN on silicon qSH mode (same as the devices described in this manuscript), Y-Z LN rayleigh mode, and X-cut LN suspended plate. Acoustic propagation losses typically scale as $f^2$, making it a major obstacle at high frequencies. While at low frequencies, the high $k^2 = 13.8$ of the q-SH mode help overcome the relatively large propagation losses of 35 dB/cm at 1 GHz. The Y-Z rayleigh mode has a poorer $k^2 = 4.9\%$, but the lower propagation loss of 1.1 dB/cm at 1 GHz make it a more suitable candidate for scaling to and past 20 GHz. Finally, for ultra high frequencies, both of these properties become important. The ultra-low losses found in suspended plate LN make it an excellent candidate for ultra high frequency operation. Additionally, this platform supports an exceptionally high $k^2 = 35\%$ acoustic mode, another desirable property.

\begin{figure*}[htbp]
    \centering
    \includegraphics[width = \linewidth]{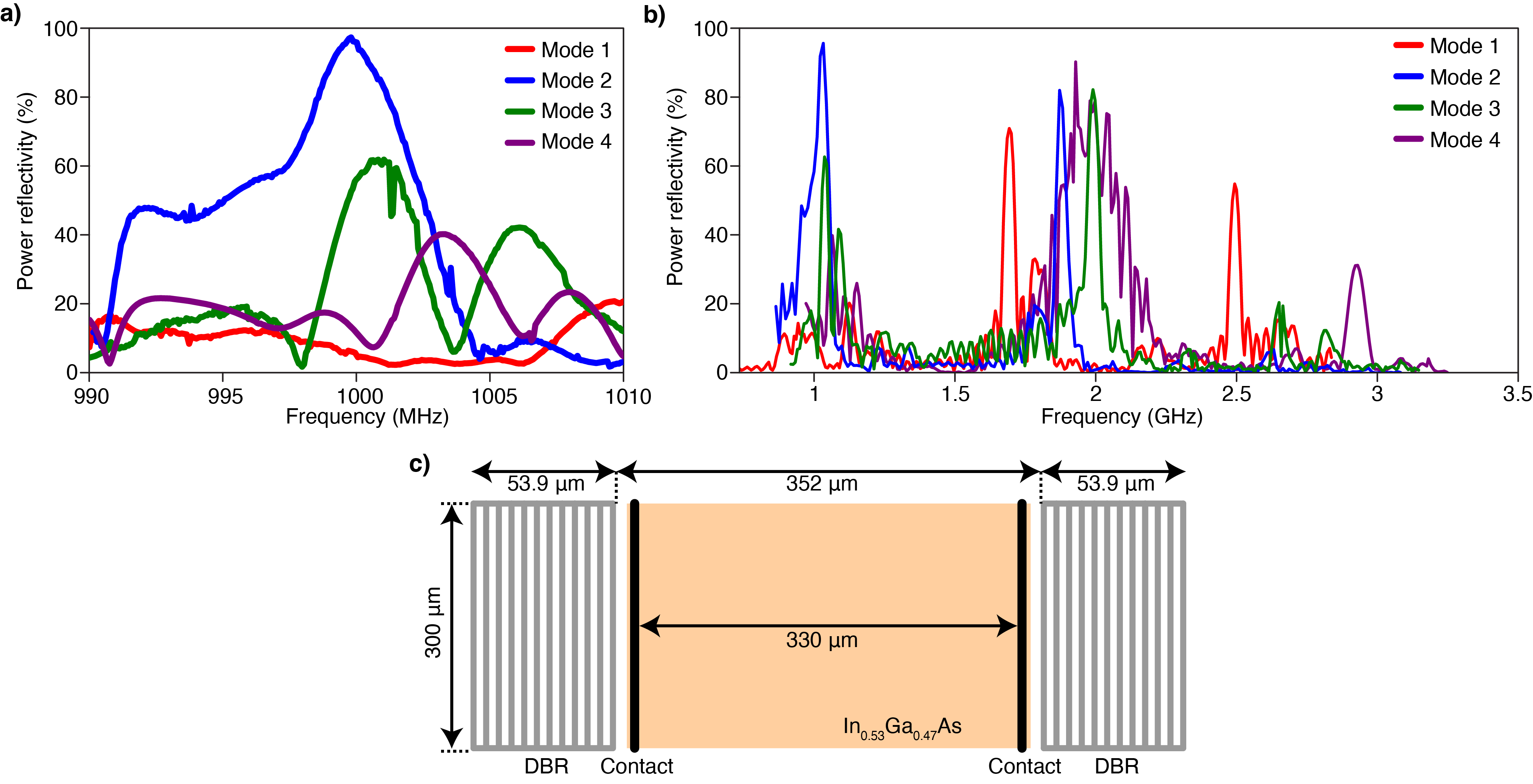}
    \caption{FEM simulation of the acoustic DBRs with 25 periods. a) Reflectance of modes around 1 GHz. Main mode of interest (q-SH0) has highest reflectivity. b) Reflectance of modes at higher frequencies. Main mode of interest has highest reflectivity over the entire spectrum.}
    \label{fig:supp: comsol mirrors}
\end{figure*}

\subsection{Resonator design}
As stated in the main text, our distributed Bragg reflectors (DBRs) consisted of a series metallic fingers with $1.1$\si{\micro\meter} lines and spaces. These metal fingers consisted of a 5\si{\nano\meter} titanium adhesion layer, a 100 \si{\nano\meter} thick aluminum layer, and a thin ($<5$\si{\nano\meter}) gold layer to prevent oxidation. The fingers were shorted together using a thin metallic strip on either end of the mirror. The device described in the main text used $N_f = 25$ total fingers for each mirror, which we found experimentally to provide the best trade-off between acoustic confinement and loss.

The reflectivity and bandwidth of a DBR depends on many design parameters, including the number of fingers, the thickness to wavelength ratio, the electromechanical coupling coefficient, and the wavenumber of the mode \cite{morgan2010surface}. Assuming no propagation loss, no scattering loss, uniform $k^2$, and no dispersion, basic theory predicts that DBRs should be reflective at every harmonic of it's fundamental reflection band. However, in a real device, all these effects vary as a function of frequency, making the true reflection band practically infeasible to analytically solve.

Instead, we performed an FEM simulation using COMSOL multiphysics to model the frequency response of our mirrors as a function of frequency for all SAW modes found using an eigenfrequency simulation, results are shown in supplemental figure \ref{fig:supp: comsol mirrors}. Around 1 GHz, we find that the simulation predicts the mode of interest to have a peak power reflection coefficient of .9433. We also predict that other SAW modes, will have non-trivial reflectance. In particular, there is a 2nd-order quasi-shear horizontal mode with slightly faster velocity ($\approx4742$\si{\meter/\second}) that has significant reflectivity, shown as the yellow curve in \ref{fig:supp: comsol mirrors}a. This mode was simulated to have an electromechanical coupling coefficient of $8.9\%$. While such a mode may pose complications to a traditional SAW resonator, the combination of significantly higher losses and decreased gain from reduced $k^2$ means this mode is unable to achieve self-oscillation before the intended q-SH0 mode oscillates.

Along with modes in the fundamental band, higher-order modes may also compete for gain. The same FEM simulation performed over a significantly larger frequency span is depicted in supplemental figure \ref{fig:supp: comsol mirrors}b, showing that there is significant reflectance of modes at 2 GHz. While three modes have significant reflectance ($>0.75$), all three have significantly lower simulated $k^2$ coefficients, and previous experimental results show that propagation loss is significantly higher at these frequencies.

The device described in the main text is pictured in supplemental figure \ref{fig:supp: comsol mirrors}c. It had a reflector-to-reflector length of 352 \si{\micro\meter}, with close to a 100 $\%$ fill fraction for the epitaxial semiconductor layer. The final contact-to-contact length was 330 \si{\micro\meter}. The mirrors each totaled 53.9 \si{\micro\meter}. The apertures were 300 \si{\micro\meter}.

\subsection{Insertion loss compensation}
\begin{figure*}[htbp]
  \centering
  \includegraphics[width=\linewidth]{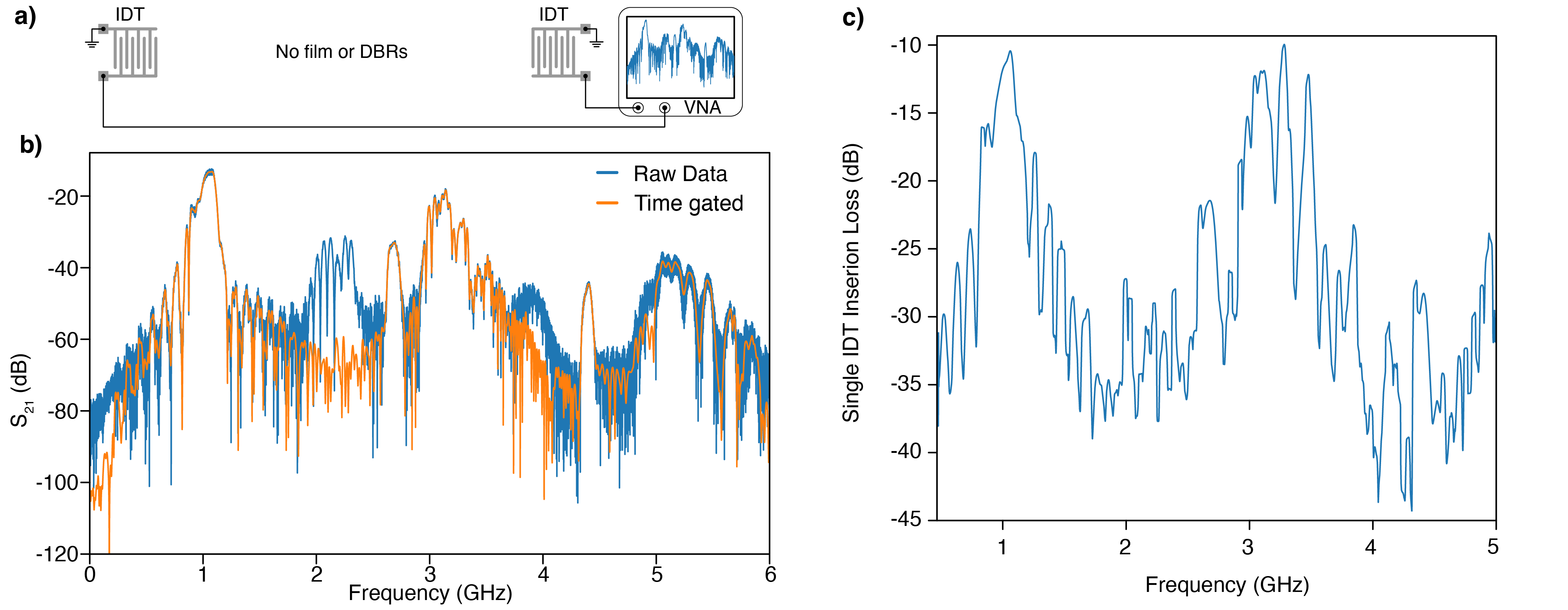}
  \caption{
    a) Delay line measurement before and after time gating. Note the resonance at 2.1 GHz disappearing, indicating it was a bulk mode. 
    b) Calculated insertion loss of a single IDT, including losses from aperture mismatch between the device and the IDT.
  }
  \label{fig:supp: IL}
\end{figure*}
How efficiently an IDT can transduce a given acoustic mode depends on many different factors, and is greatly influenced by the physical dimensions of the transducer. In this work, the IDTs for all devices had $10$ finger pairs, with finger widths of 1\si{\micro\meter} and an aperture of 100 \si{\micro \meter}, which we find to be well impedance matched around 1 GHz. While transduction is reasonably efficient in the band they are designed for, higher frequencies have poorer performance, particularly even harmonics of the fundamental. To calibrate this frequency response, acoustic delay lines consisting of two IDTs separated by lengths ranging from 400 \si{\micro\meter} to 800 \si{\micro\meter} were fabricated on the same wafer as the SAW-PLs. The S-parameters of these delay lines were measured using a VNA, shown in supplemental figure \ref{fig:supp: IL}a. These frequency sweeps were then post processed to calibrate out two unwanted responses: triple transit echo and bulk mode reflections. 

IDTs are inherently reflective, so while the intended acoustic signal will traverse the delay line once, unwanted signals will travel multiple times. IDTs also may excite bulk acoustic modes, which travel into the substrate, reflect off the bottom surface, and return to the top surface. The contributions to the frequency response due to these signals may be filtered out by applying filters in the time domain, detailed in \cite{hackett2023non}, shown in supplemental fig \ref{fig:supp: IL}b. All IDT designs are identical, however, fabrication imperfections can cause slight variations between IDTs. 

Once the delay line data was time gated, the propagation loss as a function of frequency was estimated by averaging the data from each length of delay line and fitting the best propagation loss to the data. This propagation loss was individually removed from the insertion loss of each delay line, after which the mean of the insertion loss over all delay lines was taken to find the final insertion loss. After smoothing, the square root of this data (one half in dB scale) was then used as the single IDT insertion loss.

The devices detailed in the main text had apertures of 300 \si{\micro\meter}, while the IDTs used for readout only had apertures of 100 \si{\micro\meter}, meaning a significant percentage of the acoustic output power was unable to be transduced. The IDTs were fabricated 5\si{\micro\meter} from the SAW-PLs, close enough so that diffraction effects can be ignored. While the exact spatial profile of the output wavefront is unknown, it is safe to assume that it is close to plane waves across the entire aperture. Thus, we estimate that approximately 1/3 of the acoustic power is transduced, resulting in an additional -4.77 dB of loss when calculating the acoustic power. Supplemental Fig. \ref{fig:supp: IL}c depicts the final transduction efficiency as a function of frequency.

While the device studied in this work text used metallic electrode based DBRs, acoustic mirrors may be constructed in many ways. Etching shallow trenches periodically has proven to yield high Q acoustic resonators \cite{shao2019phononic}. Additionally, as discussed in greater detail in the supplemental section of the main text, structures that do not rely on a periodic series of sub-wavelength features can be used to guide an acoustic mode to create a traveling-wave resonator. 

\begin{figure*}[htbp]
    \centering
    \includegraphics[width = \linewidth]{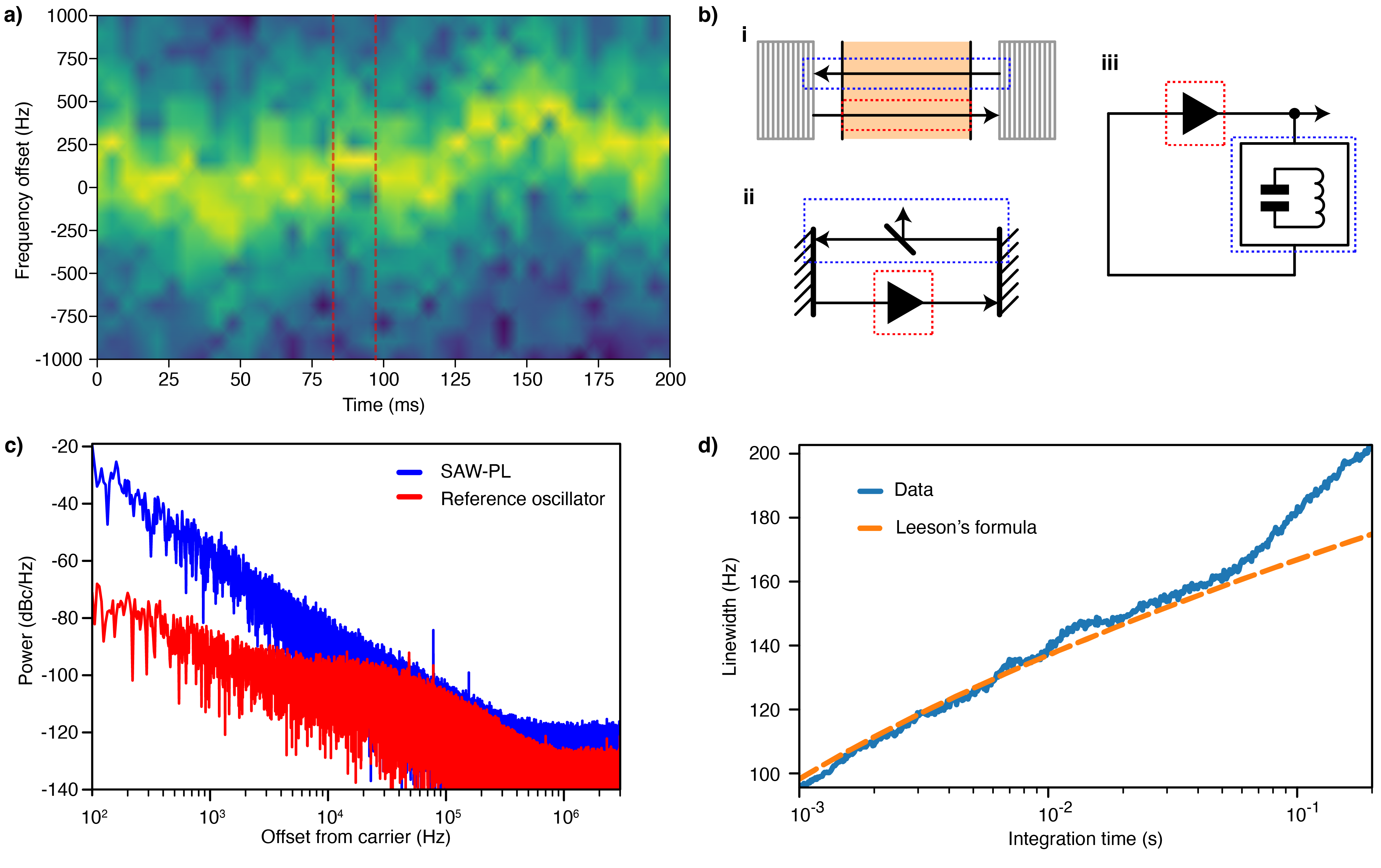}
    \caption{\textbf{a)} Spectrogram of the linewidth data \textbf{b)} The SAW-PL as a traditional electronic oscillator. Blue boxes indicate the lossy delay line/resonator element of the oscillator\textbf{i} Path of a phonon as it traverses the SAW-PL. \textbf{ii} Optical cavity analogy. \textbf{iii} the traditional representation of an electronic oscillator with an RC circuit as the resonator. \textbf{c)} Calculated SAW-PL phase noise against a low noise reference oscillator
    \textbf{c)} Linewidth calculated from the phase noise. Reprint of Figure 3f from the main text.}
    \label{fig:supp: linewidth}
\end{figure*}

\subsection{Linewidth and phase noise analysis}
As stated in the main text, there appears to be a slow, random walk-like jitter to the center frequency of the oscillator. This can be seen from a spectrogram of the full 200 ms data set, plotted in supplemental figure \ref{fig:supp: linewidth}a. The red lines indicate the most stable section, where the data used in the main text was taken from. One can see that in short intervals, the linewidth appears narrow, often limited by the 100 Hz resolution of the spectrogram FFT. However, across the entire file, the center frequency jitters over a range of about 750 Hz. We identify and analyze three primary sources of noise: Electrical noise from the DC power supply, thermal noise, and vibrational noise. 

The power supply used to apply the DC bias was a Keithley 2450 SMU, which according to it's specifications has a $<10$ Hz RMS voltage noise of 1 mV for the 200 V range required to operate at 36 V. Simply multiplying the tuning rate found in the main text by this RMS noise yields an expected noise of 60 Hz. While this is a strong candidate for the limit and could have contributed to the specific set of data presented here, this was ruled out as the sole limiting factor after repeating the experiment using several 9 V batteries instead of the SMU, which yielded similar measurements.

Another source of noise are the temperature fluctuations of the medium, which alter the acoustic velocity. In Y-cut lithium niobate, the temperature change acoustic velocity coefficient for the Rayleigh mode $\Delta v_a \approx 87$ppm \cite{ward1990temperature}. While data for the q-SH0 mode is not available, it is reasonable to assume it is of this order of magnitude. By using a mode volume of $0.1 \text{mm}^2\times 5$\si{\micro\meter}, we estimate the heat capacity of the Lithium Niobate to be $C_{LN} = 1.4\times10^{-6}$\si{\joule \per \kelvin}. From this, we find the RMS temperature fluctuations $\Delta T_{RMS} = \sqrt{k_B T^2/C_{LN}}$ to be $\approx1$\si{\micro\kelvin}. Thus, we expect the resonant frequency to deviate by $0.1$\si{Hz} on average due to thermal fluctuations. While this mechanism does not limit the devices presented in this paper, this could be a barrier to future devices, particularly ones implementing ring resonator topologies. This could be addressed, however, by selecting a different cut or material. 64 degree Y cut Lithium Niobate and 42 degree Y cut of Lithium Tantalate have temperature change coefficients of $\approx10$ ppm and $\approx1$ ppm, respectively, decreasing this limit down to the mHz level.

Finally, we estimate the impact of vibrational noise. The fractional change in acoustic velocity in response to stress is up to $1.74\times10^{-11}$ Pa$^{-1}$ \cite{cho1987nonlinear}. In lithium niobate, this translates to a fractional change in velocity of 2.96 / strain. As the resonant frequency of a resonator is proportional to velocity, a low frequency vibration of magnitude about 33 nano strain would result in a frequency shift of 100 Hz. As a rough estimate, Newport gives a relative motion of 0.2 nm between any two points. A device 400 \si{\micro\meter} long could thus have a strain of 500 nano strain, more than enough to shift the linewidth hundreds of Hz. While this is a plausible noise source, this may be addressed in future devices by selecting a material that is less sensitive to strain-induced velocity perturbations.

To estimate the intrinsic linewidth, we use the Schawlow-Townes equation for the linewidth of a maser, given by
\begin{align}
    \delta \nu = 2\pi k_B T (\Delta \nu)^2 / P_{out}
\end{align}
where $\Delta \nu$ is the FWHM of the resonator, $k_B$ is the Boltzmann constant, $T$ is the temperature in Kelvin, and $P_{out}$ is the acoustic output power of the device \cite{schawlow1958infrared}. Due to coupling between the real and imaginary parts of the propagation constant, this is broadened by the Henry factor, $\alpha$, as $\delta \nu ' = (1+\alpha^2) \delta \nu$ \cite{henry1982theory}. This factor can be calculated from the ratio of partial derivatives of these components with respect to carrier concentration:
\begin{align}
    \alpha = \frac{\Delta\beta}{\Delta g} = \frac{d(\beta_{ae}^+ +\beta_{ae}^-)/dN}{d(g_{ae}^++g_{ae}^-)/dN} 
\end{align}
where $+, -$ superscripts indicate propagation with and against the bias, respectively. In a typical maser or laser cavity, the linewidth $\Delta \nu$ is simply the FWHM of the resonator. However, in a SAW-PL with Fabry-perot topology, the acoustoelectric losses must be accounted for. These losses dominate the the intrinsic loss, significantly broadening the linewidth. By simultaneously measuring the oscillation signal from the front and back ports of the SAW-PL, we estimate the acoustoelectric loss to be -20.8 dB. By defining the quality factor as the ratio of the energy conserved to the energy lost per cycle, the effective Q factor may be estimated as $Q_{\text{eff}} = \pi\nu/v_a\alpha_l$, where $\alpha_l$ is the exponential loss factor \cite{gokhale2014phonon}. We estimate this as $Q_{\text{eff}} = 89$, resulting in a FWHM of 10.2 MHz, almost an order of magnitude larger than the passive resonator characterized in the main text. 

As discussed extensively in the main text, a ring resonator topology does not suffer this acoustoelectric loss. To estimate the improvement in intrinsic linewidth, we compute $\delta \nu$ using the passive resonator linewidth of $1.2$ MHz, which gives a limit of $0.15$ mHz. This, naturally, will be broadened by the Henry factor. However, the definition of the Henry factor no longer has the $\beta^-$ or $g^-$ parts. Using $\alpha = \frac{d\beta^+/dN}{dg_{ae}^+/dN}$, the alpha factor decreases in magnitude to 3.0, yielding a broadened linewidth of 1.5 mHz, or a factor of 500 improvement over the Fabry-Perot topology.

We can think of the SAW-PL as a conventional electronic oscillator, allowing the use of Leeson's formula to model the phase noise. Supplemental figure \ref{fig:supp: linewidth}bi illustrates this analogy. Consider the round trip path of the acoustic field in the SAW-PL, starting on the left side of the amplifier. First, it travels through the amplifier, becoming amplified. Then, it then exits the amplifier, subsequently being reflected by the right mirror. The field then travels backward through the amplifier, losing amplitude due to attenuation. Finally, it reflects of the left mirror to complete the round trip. The loss from the two mirrors and the acoustoelectric attenuation may be thought of as a composite lossy channel, illustrated by a beam splitter in supplemental figure \ref{fig:supp: linewidth}bii. Furthermore, this lossy channel has some effective Q factor, similar to how an LC tank circuit has a loss rate in a traditional electronic oscillator, shown in supplemental figure \ref{fig:supp: linewidth}biii.

Leeson's formula is given by
\begin{equation}
    L(f_m) = 10\log_{10}\bigg[\frac{Fk_BT}{2P_0}\bigg(\bigg(\frac{f_0}{2Q_{\text{eff}}f_m}+ 1 \bigg)^2 \bigg(\frac{f_c}{f_m}+1 \bigg) \bigg]
    \label{eq:supp: Leeson's supp}
\end{equation}
where $f_0$ is the center frequency, $f_m$ is the offset from the center frequency, $f_c$ is the corner frequency, F is the noise factor, $k_B$ is the Boltzmann constant, T is the temperature, $P_0$ is the acoustic power incident on the internal amplifier, and Q is the quality factor of the resonator \cite{rhea1990oscillator}. The Q factor of a resonator or delay line is defined as the ratio of the energy conserved to the energy lost per cycle of operation. For a mechanical resonator, this is given by
\begin{align}
    Q = \frac{\pi\nu}{v_a\alpha}
\end{align}
where $\nu$ is the frequency of the mode, $v_a$ is the acoustic velocity, and $\alpha$ is the total propagation loss coefficient. In our model, we split the loss into two sources, the passive resonator losses (mirror and propagation losses), and active loss (from backwards acoustoelectric loss). By simultaneously measuring the front and back outputs of the SAW-PL, we estimate there is about -17.8 dB of acoustoelectric loss each pass. By combining this loss with the passive resonator losses, we find a total effective Q factor of about 96. Similarly, we can use this acoustoelectric loss and mirror losses in tandem with the acoustic output power of -6.1 dBm to estimate the acoustic power incident on the amplifier to be 2.4 \si{\micro\watt}. Finally, we use the previously characterized noise figure for the amplifier of 2.8 dB \cite{hackett2023non}. We can then fit the flicker corner to our data, finding $f_c = 50$ kHz.

Phase noise measurements are susceptible to timing jitter of the instruments used to digitize the signal. To verify that our oscilloscope was not limiting the measurement, we characterized the phase noise of a known low noise RF source (Berkeley Nucleonics model 855B), rated for a maximum of -76 dBc/Hz at a 10 Hz offset from the carrier at 1 GHz. Using the same technique used for calculating the SAW-PL's phase noise, described in the methods section, we calculate the phase noise, shown in supplemental figure \ref{fig:supp: linewidth}c. As one can see, the oscilloscope's jitter has substantial jitter, limiting the measurement of the reference oscillator. However, at offsets below 10 kHz, the phase noise of our device is substantially above this limit, and thus does not limit the measurement at present time. Future measurements of improved oscillators likely will be below this noise floor and require more precise characterization.

The linewidth may be calculated by integrating the frequency noise spectral density, $S_f(f)$, over a defined integration time $T$:
\begin{align}
    \Delta \nu (T) = \sqrt{\int_{1/T}^{f_{\text{max}}} S_f(f) df}
\end{align}
where $f_{\text{max}}$ is the upper frequency bound of the integration. The frequency noise spectral density is related to the phase noise spectral density, $S_{\phi}(f)$, by $S_f(f) = f^2S_{\phi}(f)$. We set $f_{\text{max}} = 10$ kHz, as this is sufficient to capture the main low-frequency contributions to the linewidth while being far away from the regime of the oscilloscope jitter adding noise. We then find the linewidth both using the experimental data and from Leeson's formula, shown in supplemental figure \ref{fig:supp: linewidth}d. The measured data fits the model quite well, until about the 100 ms mark. This indicates that the slow jitter phenomena described above begins to dominate at these times scales.

Not only are phase noise measurements susceptible to instrumental noise, they are also susceptible to environmental noise. Leeson's formula only accounts for internal noise sources, not external. Contributions from external sources add linearly
\begin{align}
    L_{\text{total}}(f) = 10 \log_{10}[10^{L_{\text{internal}}(f_m)/10} + 10^{L_{\text{external}}/10}] .
\end{align}
As described above, there is reason to believe there is external noise influencing the system, particularly at low ($<$10 Hz) frequencies. While external noise could obscure the true intrinsic phase noise of the device, we believe this is not the case for our system as a reasonable set of parameters calculated from other measured system parameters seems to closely follow the observed spectrum. Care should be taken in the future to ensure that environmental noise does not limit the phase noise of the device. 

\subsection{Estimation of semiconductor parameters}
The acoustoelectric gain curve strongly depends on the amplifier layer properties, namely the charge carrier mobility $\mu$ and the charge carrier density $n$. The yield of this particular wafer was unfortunately low, and none of the hall measurement devices typically used were usable to estimate the carrier mobility. Thus, we fit the available data to our models to make the best estimate of the parameters. Previous work utilizing this semiconductor fabrication process have yielded mobilities of $\mu = 2000 -4000$ \si{\centi\meter^2/\volt\second}, and carrier concentrations between $1-3\times10^{16}$ \si{\centi\meter^{-3}} \cite{hackett2021towards, hackett2023non, hackett2024giant}, so we expect the film to have similar properties.

\begin{figure*}[htbp]
    \centering
    \includegraphics[width = \linewidth]{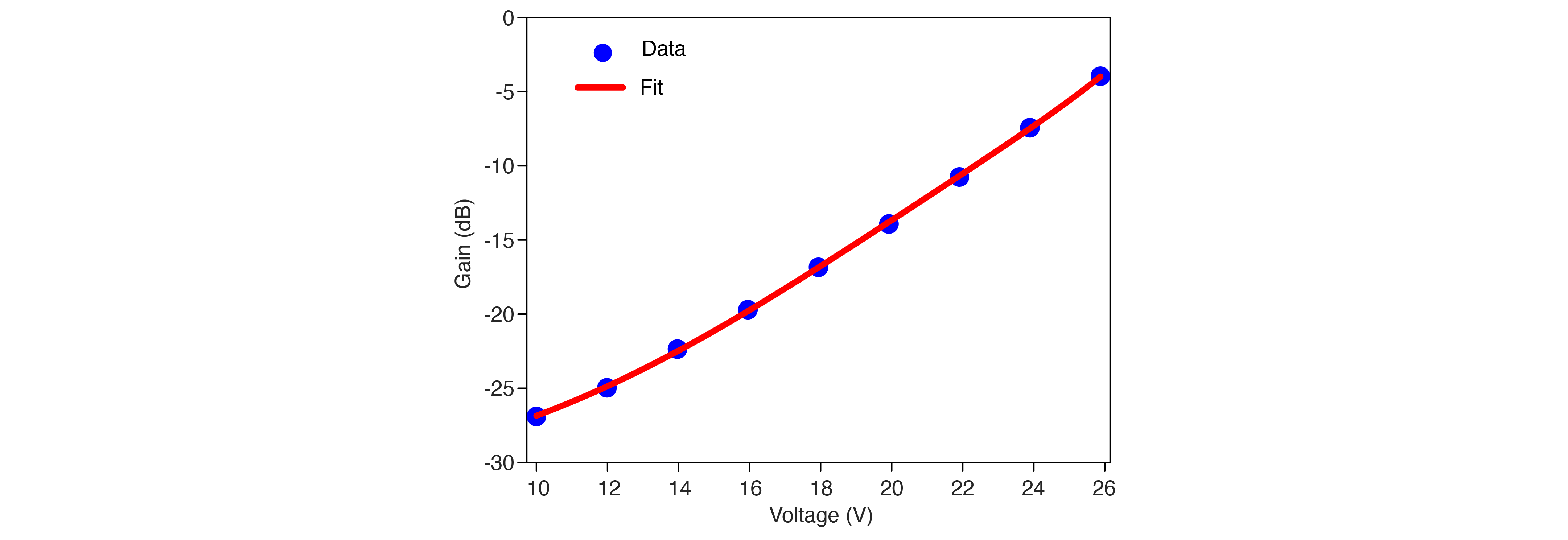}
    \caption{Estimating semiconductor parameters using the peaks of the resonant amplifier data as a function of voltage.}
    \label{fig:supp: parameter fitting}
\end{figure*}

Experimentally, we observe a threshold voltage between 30-36 V, depending on how long the device has been on and has warmed up. Based on these values, our analytical model predicts the carrier concentration is about $4\times10^{15}$ \si{\centi\meter^{-3}}, close to targeted value of $1\times10^{16}$ \si{\centi\meter^{-3}}, and a mobility of about $\mu = 4000$ \si{\centi\meter^2/\volt\second}, in line with previous results.

The rate at which the resonant frequency is tuned is also a function of voltage. The total round-trip phase accumulation is
\begin{align}
    \phi(\omega, V) = &2\beta_aL_{\text{eff}} + 4\beta'(\omega, 0)L_{\text{extra}} + \\
    &(\beta'(\omega, V) + \beta'(\omega, -V))L_{\text{amp}}
\end{align}
where $L_{\text{eff}}$ is the unperturbed effective resonator length, $L_{\text{extra}}$ is the 'extra' length of the semiconductor between the contact and the edge on either side of the amplifier that receives no bias, and $L_{\text{amp}}$ is the length of the amplifier layer. If the resonator longitudinal mode number does not change (which would result in a discontinuity that would be observed experimentally), then $\phi(\omega_1, V_1) = \phi(\omega_2, V_2)$. From this, we calculate a carrier concentration of $4\times10^{16}$\si{\centi\meter^{-3}}, with a mobility of $\mu = 4000$\si{\centi\meter^2/\volt\second}. Using the carrier concentration found from the threshold voltage, we find a tuning rate of 291 kHz/V. These values are well within a reasonable range of parameters. Furthermore, the saturation mechanism is currently unknown, and likely alters the tuning rate, making this measurement less reliable than the threshold voltage, which happens before the gain saturates.

The last method we employed to estimate the mobility and carrier density was to fit the sub-threshold resonant amplifier data. Using the circulating field approach \cite{ismail2016fabry}, one can modify the airy distribution of a Fabry-Perot interferometer to include the contributions from the forward gain $\alpha_f$ and the backward loss $\alpha_b$ to be
\begin{align}
    \frac{E_{\text{trans}}}{E_{\text{inc}}} = \frac{t_1 t_2 e^{(\alpha_f-\alpha_{\text{prop}}) L}}{1-r_1r_2 e^{-i2\phi}e^{(\alpha_f + \alpha_b - 2\alpha_{\text{prop}})L}}
\end{align}
where $t_{1,2}$ are the field transmission coefficients of the two mirrors, $r_{1,2}$ are the field reflection coefficients of the two mirrors, $\phi$ is the total phase accumulation, $L$ is the length of the resonator, and $\alpha_{\text{prop}}$ is the propagation loss. From previous work \cite{hackett2023non}, the propagation loss at 1 GHz was characterized to be $\approx 35$ dB/cm. By extracting the peak amplitude values from the resonant amplifier data as a function of voltage, we can fit the data, shown in supplemental figure \ref{fig:supp: parameter fitting}, to estimate a mobility of $\mu = 1000$ \si{\centi\meter^2/\volt\second} and $N \approx 1.5\times10^{15}$ \si{\centi\meter}$^{-3}$, again in good agreement with what is expected.

\bibliographystyle{ieeetr}
\bibliography{apssamp}